\documentclass[12pt]{iopart}

\usepackage{epsfig,graphics}
\usepackage{longtable}
\usepackage{amssymb}
\usepackage{graphicx}
\usepackage{color}
\expandafter\let\csname equation*\endcsname\relax
\expandafter\let\csname endequation*\endcsname\relax
\usepackage{amsmath}
\def \be  {\begin{equation}}
\def \ee  {\end{equation}}
\def \ee  {\end{equation}}
\def \bea {\begin{eqnarray}}
\def \eea {\end{eqnarray}}

\renewcommand{\figurename}{{\bf Fig.}}
\renewcommand{\tablename}{{\bf Tab.}}
\begin{document}
\title {Temperature-dependent nuclear partition functions and abundances in
 stellar interior}
\author{Jameel-Un Nabi$^1$ \footnote{Corresponding author}, Abdel Nasser Tawfik$^2$, Nada Ezzelarab$^2$, Ali Abas Khan$^1$}
\address{1 Faculty of Engineering Sciences, GIK Institute of Engineering Sciences and
Technology, Topi 23640, Khyber Pakhtunkhwa, Pakistan
\\ 2  Egyptian Center for Theoretical
Physics (ECTP), Modern University for Technology and Information
(MTI), Al-Mokattam, 11571 Cairo, Egypt} \eads{jameel@giki.edu.pk}

\begin{abstract}

We calculate temperature-dependent nuclear partition functions
(TDNPFs) and nuclear abundances for $728$ nuclei assuming nuclear
statistical equilibrium (NSE). The theories of stellar evolution
support NSE. Discrete nuclear energy levels have been calculated
\textit{microscopically}, using the pn-QRPA theory, up to an
excitation energy of $10$ MeV in the calculation of TDNPFs. This
feature of our paper distinguishes it from previous calculations.
Experimental data is also incorporated wherever available to ensure
reliability of our results. Beyond 10 MeV we employ simple Fermi gas
model and perform integration over the nuclear level densities to
approximate the TDNPFs. We calculate nuclidic abundances, using the
Saha equation, as a function of three parameters: stellar density,
stellar temperature and lepton-to-baryon content of stellar matter.
All these physical parameters are considered to be extremely
important in stellar interior. Results obtained in this paper show
that the equilibrium configuration of nuclei remains unaltered by
increasing stellar density (only calculated nuclear abundances
increases by roughly same order of magnitude). Increasing the
stellar temperature smooths the equilibrium configuration showing
peaks at neutron-number magic nuclei.
\end{abstract}
 \maketitle


\section{Introduction}

Temperature-dependent nuclear partition functions (TDNPFs) play a
key role in many astrophysical phenomena, e.g. nuclear abundance
calculations beyond silicon burning phases of massive stars
\cite{Hix96} and determination of equation of state during
gravitational collapse \cite{Ful82}. An accurate picture of the
nuclear level density is a pre-requisite for the calculation of
TDNPFs. The largest uncertainty in the calculation of nuclear
abundances originates from the treatment of nuclear partition
function \cite{Liu07}. Considerable improvements have been done in
the recent past for a better description of nuclear level density.
Collective effects in the level density were incorporated for a
better agreement between observed and predicted energy levels
\cite{Can14}. It was Rauscher and Thielemann \cite{Rau00} who used
the Hauser-Feshbach formalism to perform a statistical calculation
of TDNPFs for around 5000 nuclei reaching neutron and proton
driplines, using two different mass models, covering temperatures up
to 10 GK. Later Rauscher \cite{Rau03} extended his two sets of
calculation to include high-temperature corrections and performed
calculation of TDNPFs up to a stellar temperature of 275 GK. It was
also highlighted that for calculations involving isotopic abundances
in presupernova cores, the low-lying nuclear energy levels need to
be treated as discrete and a summation over these states provided a
far better description of nuclear partition functions \cite{Dim02}.
Authors in Ref. \cite{Cow91} laid much emphasis on low-lying
discrete nuclear energy states for calculation of reaction rates in
the $r$-process. Similarly Hix and Thielemann \cite{Hix96}
recommended inclusion of discrete low-lying nuclear energy levels
for an accurate determination of partition functions leading to a
better understanding of the physics of quasi-equilibrium processes
during the silicon burning phases of massive stars. Authors in
\cite{Dim02} calculated deviations up to 50$\%$ in the calculated
nuclear abundances with and without inclusion of low-lying discrete
energy levels in the calculation of nuclear partition functions. In
this paper we undertake a novel treatment of calculation of TDNPFs.
We calculate low-lying discrete energy levels up to 10 MeV
excitation energies in a microscopic fashion using the pn-QRPA
model. Beyond 10 MeV a simple level density function was assumed and
integration was performed up to 25 MeV excitation energy. We also
inserted all available measured energy levels along with their spins
in our calculation. It is hoped that this recipe would lead to a
more realistic description of TDNPFs and a better understanding of
the stellar matter beyond silicon burning phases of massive stars.

Any theory of nucleosynthesis addressing the problems of determining
isotopic abundances of chemical elements and evolution of Galaxy is
a challenging task and involves input from all branches of
astrophysics and physics in general. To date there exists no
completely unique and clear picture of nucleosynthesis and
associated theory explaining evolution of Galaxy.  The complex
dynamics of stellar evolution is dictated at large by nuclear
reactions occurring in the stellar core. The nuclear reactions
occurring in stellar environment have the most prominent influence
on element synthesis from nuclear physics point of view. At stellar
temperatures (between 10$^{9}$ K and 10$^{11}$ K) and at suitable
densities, all types of nuclear processes (e.g. $(\alpha, \gamma)$,
$(p, \gamma)$, $(n, \gamma)$, $(\gamma, \alpha)$, $(\gamma, p)$,
$(\gamma, n)$, $(p, n)$) occur profusely.  These nuclear reactions
become so abundant that an approximate statistical equilibrium is
established between the concentrations of nuclide. The nuclear
statistical equilibrium (NSE) is then said to be established. Once
NSE is established, any nucleus can be transformed into any other
nucleus under consideration, and the radiation in the assembly must
be in equilibrium with the matter. The thermodynamic conditions
spanned over several $100$ ms time interval post bounce in
core-collapse supernova simulation have been analyzed \cite{Fis14},
where the density range spanned around $10$ orders of magnitude, the
temperature around three orders and the lepton fraction more than
one order of magnitude.  It is obvious that the construction of NSE,
which is valid over such a huge thermodynamic domain is far from
being a trivial task. For example, at high densities various types
of interactions among the constituents can not be neglected. At high
temperatures the eventual inconsistencies between cluster and free
nucleon definitions should be taken into account. Detailed
discussion may be seen in the seminal paper of Hoyle \cite{Hoy46}
which showed that the radiation must be in statistical equilibrium
with the matter to a high degree of validity at temperatures of
several billion degrees. Assuming NSE, it is possible to approximate
the nucleosynthesis of elements by relatively simple calculations
independent of specific supernova models \cite{Har85}. This indeed
provides more freedom to analyze the nucleosynthesis problem as a
first-order approach. In the present paper a model-independent
approach is adopted. Theories of stellar evolution suggest that in
stellar interior interaction between radiation and matter is so
intense that approximate statistical equilibrium is eventually
reached between the concentrations of different nuclei. Under such
conditions the most strongly bound nuclei (iron-region nuclei) tend
to be most abundant. In this context the equilibrium theory may be
able to explain the observed nuclear abundances in the iron region.
For related discussions we refer to the classical paper by Clifford
and Tayler \cite{Cli63}.

Later studies of nuclear reaction rates indicate that all known
nuclei may be transformed into any other nucleus by nuclear
reactions on a time scale ranging from a day to millions of years,
depending on the temperature of the core \cite{Tsu65}. It was
discussed in \cite{Tsu65} that the concerned nuclear reactions
involve any combination of projectiles in and projectiles out for
four types of projectiles: neutrons, protons, alpha particles, and
photons. In this paper we shall be concerned with determination of
the actual abundances of nuclei in statistical equilibrium as a
function of stellar temperatures and densities (which in turn
determine the Fermi energy of the associated electrons and
positrons). These conditions may apply to the presupernova phases to
late stages of stellar evolution immediately preceding a supernova
explosion, as well as to the remnant of the explosion, which may be
a neutron star. It is necessary to specify some relation between the
total number of neutrons and the total number of protons in the
system (including contributions from all nuclei considered in the
ensemble). Further the condition of $\beta$-decay steady state is
also a convenient condition to impose. For many years it has been
recognized that the weak interaction, specially nuclear beta-decay
and electron capture, is important during the late stages of stellar
evolution. Recently it has become clear that more work needs to be
done on these rates \cite{Ful80, Auf94, Oda94, Nab99, Bra00, Cow04,
Nab04, Gao11, Nab12}. It is useful to start with the most important
nuclei and proceed to less important ones as time and resonances
permit. Unfortunately, the nuclei which cause the largest change in
the lepton-to-baryon fraction ($Y_e$) of stellar matter are neither
the most abundant ones nor the ones with the strongest rates, but a
combination of the two. In fact, the most abundant nuclei tend to
have small rates (because they are more stable) and the most
reactive nuclei tend to be present in small quantities. Work by
Cameron \cite{Cam73} is perhaps the  standard abundance distribution
to which theoretical predictions are usually compared.

In this paper we report calculation of TDNPFs and nuclear abundances
for a range of physical parameters which are thought to be important
in stellar core assuming NSE (silicon burning phases and beyond). We
calculate the nuclear abundance as function of stellar density,
stellar temperature and lepton-to-baryon fraction of stellar matter.
Our ensemble contained a total of $728$ nuclei which were selected
primarily to cover a decent number of possible inputs for
astrophysical simulations (see Table A  of Ref. \cite{Nab99a} and
Table B of Ref. \cite{Nab04} for selection of these nuclei). For
other recent discussions on nuclear abundance calculation see
\cite{Nad04, Sei08, Bot10, Odr12, Buy13}. There are at least two
main features of the present calculation that distinguishes it from
NSE calculations in the past. Firstly we had a pool of nuclei
containing a decent number of nuclear species up to mass number 100.
Then all states up to excitation energy of $10$ MeV were
microscopically calculated (along with manual insertion of all
experimental data). A state-by-state sum over discrete energy levels
was performed to calculate the nuclear level densities and TDNPFs
for all nuclei. In future, we plan to study the combined effect of
nuclear abundance and weak decay rates to compile a priority list of
nuclei which can significantly effect the time rate of change of
$Y_e$  during  presupernova evolution of massive stars.

The present paper is organized as follows. Section~2 briefly
describes the formalism which was adopted to calculate the TDNPFs
and nuclear abundances of all nuclei present in our pool. Section~3
discusses our results and also compares our calculation with
previous calculations. We finally conclude our findings in
Section~4.

\section{Formalism}
We start from the basic assumption stated earlier that nuclei may be
treated as ideal, non-degenerate and non-relativistic gas. The
TDNPFs are given by \cite{Fow67}:
\begin{align}\label{ld}
\omega(A,Z,T)=\sum_{\mu = 0}^{\mu_{m}}(2
J^{\mu}+1)\exp[{-E^{\mu}/kT}]+\int_{E^{\mu_{m}}}^{E^{max}}\sum_{J^{\mu},
\pi^{\mu}}(2 J^{\mu}+1)\exp(-\epsilon/kT)\rho(\epsilon,
J^{\mu},\pi^{\mu})d\epsilon,
\end{align}
with $\mu_{m}$ being the label of last included experimentally known
and/or theoretically calculated energy state.  In Eq.~(\ref{ld}),
$\rho$ is the nuclear level density and other symbols have their
usual meanings. Above the last discrete state  an integration was
performed over the nuclear level density. For all the $728$ nuclei
considered in this study, the theoretical energy eigenvalues for
each nucleus were calculated within the proton-neutron quasiparticle
random phase approximation (pn-QRPA) formalism \cite{Mut92, Nab99,
Nab04} up to $10$ MeV. This is a special feature of our project
where we calculate excitation energy up to $10$ MeV in each nucleus
in a purely microscopic fashion and distinguishes this calculation
from previous ones (e.g. \cite{Hem10, Ish03, Auf94, Tsu65}). We
incorporated the latest experimental data into our calculations for
increased reliability, wherever possible. The calculated excitation
energy was replaced with an experimental one when they were within
$0.5$ MeV of each other (missing measured states were inserted)
along with their $2J+1$ values. For a particular nucleus, measured
(inserted) states with missing angular momentum were assigned the
biggest angular momentum in the list. For cases where more than one
value of $J$ was reported, we selected the highest value.

NSE is applicable roughly from $0.1$ to $100$ MeV. For the
temperature range considered in this calculation ($\sim$ 10$^{7}$--
10$^{10}$ K) excitation energies in the range $20 - 25$ MeV is safe
to achieve convergence in nuclear partition functions \cite{Rau00}.
To save space we do not describe the formalism of solving the
pn-QRPA equations here. Interested reader can see the formalism in
Ref. \cite{Mut92, Nab04}. We performed the summation to last
known/calculated state. To include contribution from excited states
beyond $10$ MeV, integration was performed.  Assuming a uniform
Fermi gas, we estimated the level density (using saddle point
approximation yields and some additional simplifications) as
\cite{Boh69}
\begin{equation}\label{lda}
\rho(A,E) = \frac{\exp(2\sqrt{a E})}{4\sqrt{3}E},
\end{equation}
where $a = \frac{\pi^{2}g}{6}$ and the density of single particle
states for a nucleus having $A$ fermions is $g =
\frac{3A}{2\epsilon_{f}}$ where $\epsilon_{f}$ is the Fermi energy.
Of course a more sophisticated treatment for calculation of the
nuclear level density  (e.g. back-shifted, pairing corrections,
inclusion of spin cut-off parameters) is possible. It is to be noted
that we microscopically calculate all energy levels up to $10$ MeV
in daughter. Beyond $10$ MeV, the approximate formula used in
Eq.~(\ref{lda}) still gives us a reliable estimate of the level
density for the calculation of TDNPFs.

In order to calculate the nuclear abundance, $n(A,Z)$, we followed
the recipe given by Ref. \cite{Bur57} which is valid under the
assumption of NSE as:
\begin{equation}\label{na}
n(A,Z)= \omega(A,Z,T)\left(\frac{A M kT}{2 \pi
\hbar^{2}}\right)^{\frac{3}{2}} \left(\frac{2 \pi \hbar^{2}}{M
kT}\right)^{\frac{3A}{2}} \frac{n_{n}^{A-Z}n_{p}^{Z}}{2^{A}}\exp
\left[Q(A,Z)/kT \right],
\end{equation}
where $n_{p}$ and $n_{n}$ are the number densities of free protons
and neutrons, respectively. $ Q(A,Z)$ is the total binding energy of
the nucleus $(A,Z)$, $\omega(A,Z,T)$ are the TDNPFs of the nucleus
$(A,Z)$, $M$ is the atomic mass unit, and the other symbols have
their usual meanings. For diluted matter, Eq. (\ref{na}) is formally
correct. However at higher densities various types of interactions
should be taken into consideration . For instance, due to electron
screening, the nuclear binding energies in stellar environment
should be different than those in the vacuum \cite{Ish03}.
Consequently, driplines are shifted and the heavy nuclei become
unstable against fission under terrestrial condition. They may
become stable in supernova matter. These might be very important
under astrophysical conditions. Seeking for completeness and to
account for the electron screening, self-consistent Hartree-Fock
approximation in Wigner-Seitz cells can be utilized \cite{WS}. The
net binding energy of the ground level of the nucleus $(A,Z)$ (see
also \cite{Hem10}) is given by
\begin{equation}\label{sbe}
Q(A,Z)^{net}= Q(A,Z) -\Delta V_{f}^{Coul}(\rho_{e}) \nonumber,
\end{equation}
\begin{equation}\label{coul}
\Delta V_{f}^{Coul}(\rho_{e}) = -\frac {3}{5} \frac
{Z^{2}e^{2}}{R_{0}}\left(\frac {3}{2} \eta - \frac {1}{2}
\eta^{3}\right)
 \nonumber,
\end{equation}
where $\rho_{e}$ is the electron density, $\eta = (\frac
{n_{e}}{n_{0}}\frac{A}{Z})$, $n_{e}$ is the number density of free
electrons and we treat nuclei as homogeneous spheres with radius
$R_{0}$ of nucleons at saturation density $n_{0}$ ($R_{0} = (3A/4\pi
n_{0})^{1/3}$).
\begin{equation}\label{be}
Q(A,Z)= c^{2}[(A-Z)M_{n} + Z M_{p} - M (A,Z)],
\end{equation}
where $M_{n}$, $M_{p}$ and $M(A,Z)$ are the masses of the free
neutron, free proton, and nucleus $(A,Z)$, respectively. We now
return to solve Eq.~(\ref{na}). The stellar density, $\rho_{st}$ can
be approximated as:
\begin{equation}\label{st}
\sum_{A,Z} M(A,Z) n(A,Z) +  \sum_{A,Z}M_{p}  n_{p} + \sum_{A,Z}
M_{n} n_{n}= \rho_{st},
\end{equation}
where we have assumed $\sum m_{e}n_{e} \approx 0$. Also assuming NSE
\begin{equation}\label{ne}
n_{e} = N_{A} \rho_{st} Y_{e},
\end{equation}
where $N_{A}$ is the Avagadro constant. Equilibrium between neutrons
and protons is established provided the time scale is long enough.
The equilibrium condition is expressed by the equations \cite{Bur57}
\begin{equation}\label{nnnp}
\frac{n_{n}}{n_{p}}= \frac{n_{e}}{2}\left(\frac{2 \pi \hbar
^{2}}{m_{e}k T}\right)^{3/2}\exp\left(\frac{-0.78 (\mathtt{MeV})}{k
T (\mathtt{MeV})}\right),
\end{equation}
\begin{equation}\label{nee}
n_{e}=\sum_{A,Z} Z n(A,Z) + n_{p}.
\end{equation}
Using  Eq.~(\ref{ne}), Eq.~(\ref{nnnp}) and Eq.~(\ref{nee}) our
Eq.~(\ref{st}) becomes
\begin{equation}\label{mn}
\sum_{A,Z} M(A,Z) n(A,Z) + \sum_{A,Z} M_{p} \left[n_{e} -
\sum_{A,Z}Z n(A,Z)\right] + \sum_{A,Z} M_{n} \frac{n_{n}}{n_{p}}
\left[n_{e} - \sum_{A,Z}Z n(A,Z)\right]= \rho_{st}.
\end{equation}
Eq.~(\ref{na}) and  Eq.~(\ref{mn}) can be used iteratively to solve
for $n(A,Z)$ for a fixed value of stellar temperature $T$, stellar
density $\rho_{st}$ and lepton content of stellar matter $Y_{e}$.
%

\section{Results and comparison}
A total of $728$ nuclei was considered in our stellar pool. List of
selected nuclei can be seen from Table~\ref{ta0}. The list was
inspired by previous compilation of stellar weak rate calculations
\cite{Oda94, Nab99, Nab04}. In this section we present our
calculations of TDNPFs, nuclear level densities, mass fractions and
nuclear abundances. We also compare our results with previous
calculations.

The novelty of current calculation lies in the treatment of TDNPFs.
Recently Seitenzahl and collaborators \cite{Sei09} solved the
equations of nuclear statistical equilibrium (NSE) for 443 nuclei
for Type Ia supernova simulations. In their calculation the authors
used the TDNPFs calculation of \cite{Rau00}. Later Juodagalvis et
al. \cite{Juo10} performed calculation of electron capture rates for
around 2700 nuclei assuming conditions of NSE. These authors used
TDNPFs calculation of \cite{Rau03} for their estimates of electron
capture rates. In this section we compare our TDNPFs calculation
with those of \cite{Rau00, Rau03}. Authors in \cite{Rau00} used the
Hauser-Feshbach formalism and performed calculation of reaction
rates for targets 10$\le Z \le $83 covering a temperature range
0.01~$\times$~10$^{9}~\le~T$(K)~$\le $~10~$\times$~10$^{9}$ which
included calculation of TDNPFs. They employed two different mass
models for their calculation of TDNPFs (we refer to their
calculation as RT throughout this section). RT used the finite range
droplet model (FRDM) \cite{Moe95} and an extended Thomas-Fermi
approach with Strutinski integral (ETFSI) \cite{Pea96} to perform
their TDNPFs calculation. Later Rauscher \cite{Rau03} extended his
calculation of TDNPFs to temperature range 12 $\times$ 10$^{9} \le
T$(K) $\le $ 275 $\times$ 10$^{9}$ with 9$\le Z \le $85 applying
high-temperature corrections. Again the same two mass models were
used to calculate two sets of TDNPFs (we refer to this calculation
as TR throughout this section).

Tables~\ref{ta1}-\ref{ta8} present our calculation of TDNPFs and
comparison with previous calculations of RT and TR. Table~\ref{ta1}
and Table~\ref{ta3} show calculation of TDNPFs for selected iron and
nickel isotopes (even $Z$) in our ensemble along with comparison
with RT calculation. No calculation was done for iron isotopes using
the ETFSI mass model by RT. It is to be noted that the calculations
of TDNPFs by \cite{Rau00, Rau03} are normalized to the ground-state
spin multiplicity whereas our TDNPFs includes the ground-state spin
multiplicity factor  (see Eq.~(\ref{ld})). Correspondingly, at low
temperatures we do notice differences in the two calculations for
odd-A isotopes of iron and nickel. RT applied a statistical
description throughout the nuclear chart without relying on
experimental level density parameters in specific cases and noted in
their paper that this may lead to locally slightly larger deviations
from experiment. We incorporated experimental energy levels (along
with their spins) in calculation of TDNPFs for all nuclei considered
in our ensemble. This result in an enhancement of our calculated
TDNPFs up to an order of magnitude for odd-A iron and nickel
isotopes. As temperature increases the two calculations come in
better agreement. For even-even cases the two calculations compare
reasonably well ($J^{\pi}$ = 0$^{+}$). RT considered excitation
energies of the order of 20--30 MeV in their calculation. Our
corresponding range was 20--25 MeV. At high temperatures the TDNPFs
increases much more rapidly with temperature. This indicates the
exponential relationship between nuclear partition function and
temperature. Table~\ref{ta2} and Table~\ref{ta4} show the comparison
of our calculated TDNPFs with those by TR at high temperatures
reaching 28 $\times$ 10$^{9}$ K for iron and nickel isotopes,
respectively. For many nuclei the comparison is fair. But there are
also cases at high temperatures where TR calculation is bigger than
ours by as much as 1--5 orders of magnitude. We further notice that
the TDNPFs calculated by TR is up to an order of magnitude bigger
than ours also for heavy odd-A iron and nickel isotopes (i.e. for
neutron-rich isotopes). We attribute these differences to
incorporation of high-temperature effects in the calculation of TR.
Further the minimum excitation energy considered in the TR
calculation was 35 MeV. Nuclear partition functions for odd-odd
nuclei are bigger than odd-A nuclei. We present comparison of TDNPFs
calculations of two odd $Z$ cases (cobalt and copper isotopes) in
Tables~\ref{ta5}-\ref{ta8} at low and high stellar temperatures.
Calculations of RT and TR also show that the TDNPFs is a sensitive
function of the input mass model. All theoretical masses were taken
from the recent mass compilation of Audi et al. \cite{Aud12} in our
calculation of TDNPFs. For cases where \cite{Aud12} failed to report
masses, we used input masses from \cite{Moe95}.

\begin{figure}
\begin{center}
\includegraphics[width=6in,height=5.5in]{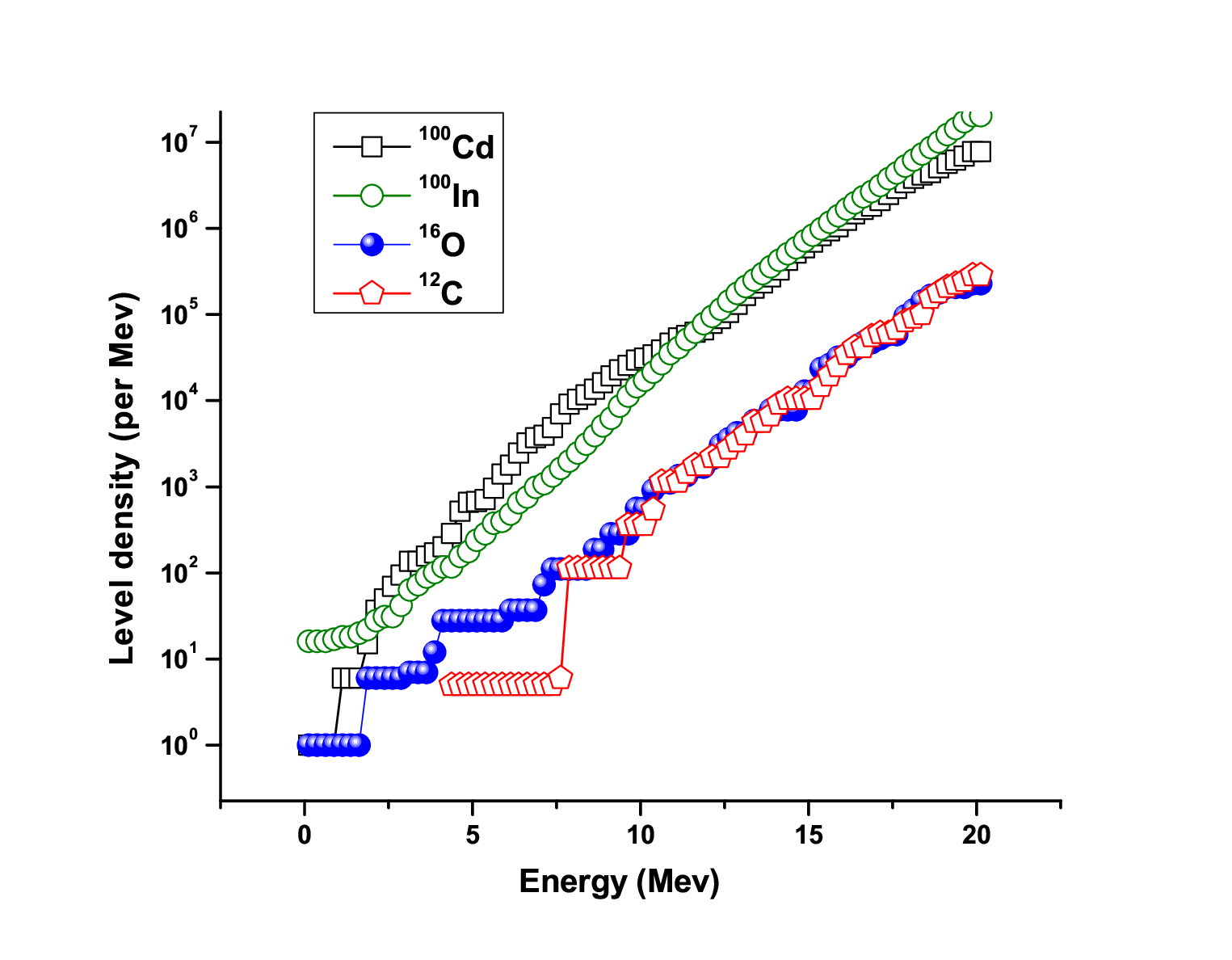} \caption{(Color
online) Calculated nuclear level densities for selected nuclei using
the pn-QRPA model.}\label{fig1}
\end{center}
\end{figure}

Figure~\ref{fig1} shows our level density calculation of $^{100}$Cd,
$^{100}$In, $^{16}$O and $^{12}$C. As expected there is an
exponential increase in calculated nuclear level densities with
increasing excitation energy. The calculations are shown up to
excitation energy of $20$ MeV. It is again stated that the pn-QRPA
model was used to microscopically calculate all energy eigenvalues
up to $10$ MeV in our calculations. Between $2$ and $11$ MeV, our
model calculate biggest level density for $^{100}$Cd. As excitation
energy increases, the level density of odd-odd nucleus $^{100}$In
becomes more than the even-even $^{100}$Cd. Figure~\ref{fig1} also
shows our level density calculation for two even-even lighter nuclei
$^{16}$O and $^{12}$Mg.

The calculation of mass fraction was initially performed with the
open source nuclear statistical equilibrium code \textit{libnuceq}
\cite{net}. We refer to this code as OSNSEC throughout this paper.
Tables~\ref{ta9}-\ref{ta12} show the comparison of our calculated
TDNPFs with those used in OSNSEC for selected iron, nickel, cobalt
and copper isotopes, respectively. The TDNPFs used in OSNSEC is in
fair agreement with our calculated TDNPFs up to stellar temperature
$T_9$ (in units of 10$^{9} K$) $ \sim 10$. Beyond $T_9 = 10$, our
calculated TDNPFs are 1-4 orders of magnitude bigger. At high
stellar temperatures the partition functions used in OSNSEC are also
orders of magnitude smaller than those calculated by \cite{Rau00,
Rau03}. Accordingly we used OSNSEC to calculate mass fractions only
for $T_9 \leq 7.33$.

Four sets of initial conditions, including stellar density
$\rho_{st}$, stellar temperature $T_9$  and lepton fraction content
of stellar matter $Y_e$, were chosen to perform the mass fraction
calculation. These sets of initial conditions were taken to be the
same as those by ~\cite{Auf94} (referred to as AFWH throughout this
paper) in order to compare the results with previous calculation.
The four set points roughly characterize the track that the central
region of a massive star follow after the silicon core burning.
OSNSEC calculated mass fractions and those calculated by AFWH, at
$\rho_{st} =5.86 \times 10^{7}$ g/cm$^{3} $, $T_9 = 3.40$  and $Y_e
= 0.47$, are shown in Table~\ref{ta13}.  It is to be noted that
authors in AFWH compiled a list of most influential weak interaction
nuclei during the presupernova evolution of massive stars. The most
influential nuclei were classified as those having the largest
product of weak rates (electron capture or $\beta$-decay)
\textit{and} mass fraction.  We did not incorporate the
weak-interaction mediated rates for the nuclei in our pool and this
would be taken as a future assignment. The OSNSEC calculation
resulted in abundant production of iron-regime nuclei (e.g.
$^{53-58}$Fe, $^{56-62}$Ni and $^{55-59}$Co). The top 30 abundant
nuclei according to OSNSEC calculation under given physical
conditions are shown in Table~\ref{ta14}. The two calculations
compare well. Table~\ref{ta15} shows the calculated mass fractions
and those calculated by AFWH for stellar density $\rho_{st} =1.45
\times 10^{8}$ g/cm$^{3} $, for core temperature $T_9 = 3.80$  and
for lepton fraction $Y_e = 0.45$.  We  show the 30 most abundant
nuclei produced in OSNSEC calculation under given conditions in
Table~\ref{ta16}. The mass fractions change as the stellar
temperature and density increases and the lepton content of stellar
matter decreases. The results are shown in Table~\ref{ta17} and
Table~\ref{ta19}. Table~\ref{ta17} gives mass abundances at
$\rho_{st} =1.06 \times 10^{9}$ g/cm$^{3} $, $T_9 = 4.93$   and $Y_e
= 0.43$. Finally Table~\ref{ta19} shows calculated mass abundances
at $\rho_{st} =4.01 \times 10^{10}$ g/cm$^{3} $, $T_9 = 7.33$  and
$Y_e = 0.41$. Table~\ref{ta18} and Table~\ref{ta20} show the top 30
abundant nuclei produced in our calculation for physical conditions
depicted in Table~\ref{ta17} and Table~\ref{ta19}, respectively.
Only for few cases the two calculations differ by roughly an order
of magnitude. AFWH incorporated only the measured values of ground
state spins wherever available. No discrete excited states were
calculated by AFWH and an integration was performed on all excited
states using a backshifted Fermi gas level density formula.

Authors in \cite{Liu07} performed  mass fraction calculation of
stellar matter for a relatively small pool of nuclei. They
essentially used the formalism employed by Rauscher \cite{Rau03} for
the calculation of TDNPFs and assumed NSE conditions to study
core-collapse supernovae. A direct comparison of OSNSEC calculated
mass fraction calculation with that of \cite{Liu07} can be seen in
Table~\ref{ta21}. The comparison of calculated mass fractions is
shown at $T = 5\times10^{9}$ K, $\rho_{st} = 1\times10^{7}$
g/cm$^{3}$ and $Y_{e} = 0.5$. It is evident from Table~\ref{ta21}
that the two calculations are in decent comparison. Small
differences are attributed to the fact that Rauscher used TDNPFs
normalized to the ground-state spin multiplicity.

We finally present our calculation of number density of free
protons, free neutrons, alpha particles and nuclear abundances for
nuclei considered in our pool of $728$ nuclei assuming conditions of
NSE and using Eq.~(\ref{na}). It is pertinent to mention that we
used our pn-QRPA calculated TDNPFs to calculate these abundances.

\begin{figure}
\begin{center}
\includegraphics[width=7.2in,height=5.in]{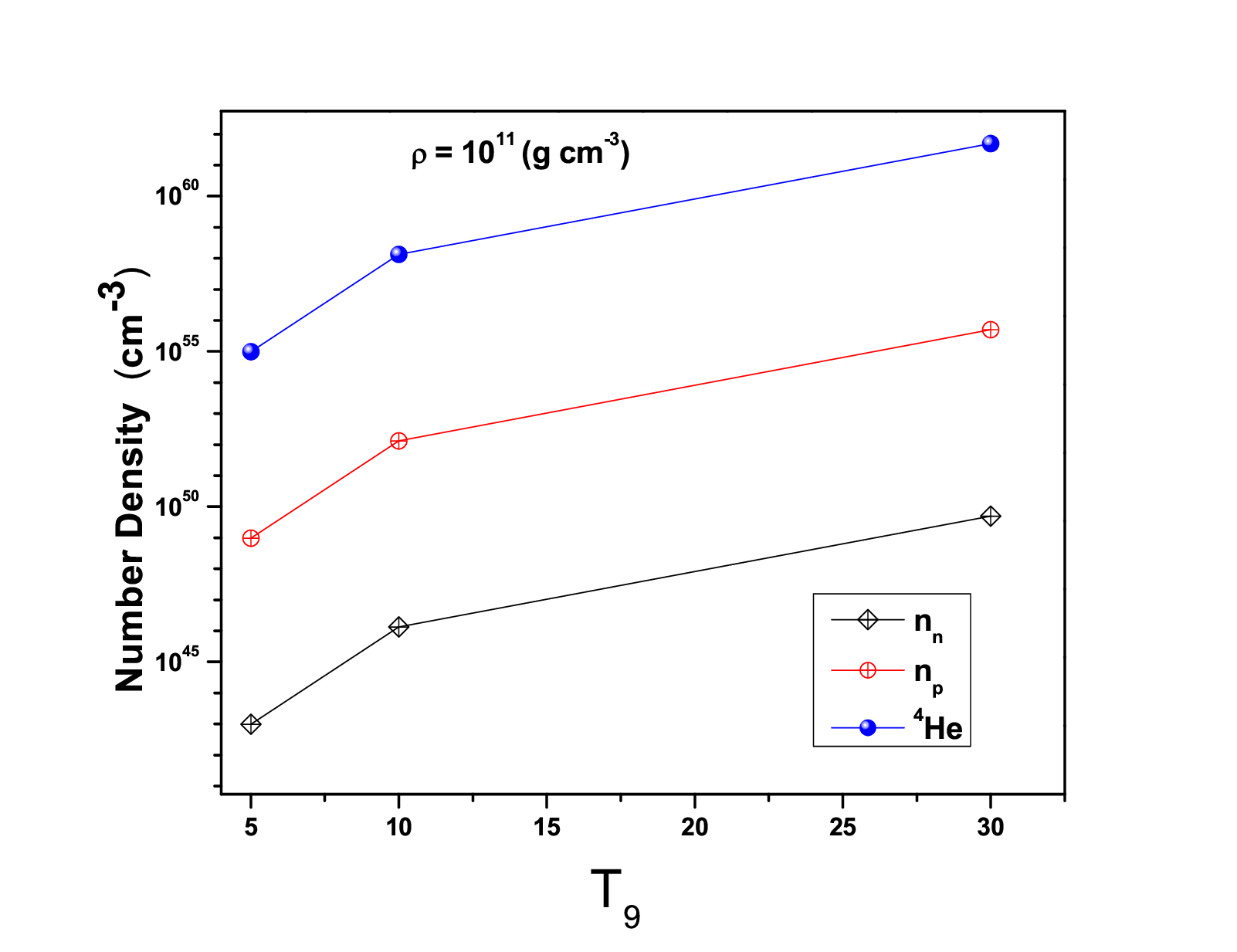}
\caption{(Color online) Number density of nucleons and alphas at
$\rho_{st} = 10^{11}$ g/cm$^{3} $ as a function of stellar
temperature.}\label{fig2}
\end{center}
\end{figure}

\begin{figure}
\begin{center}
\includegraphics[width=7.2in,height=5.in]{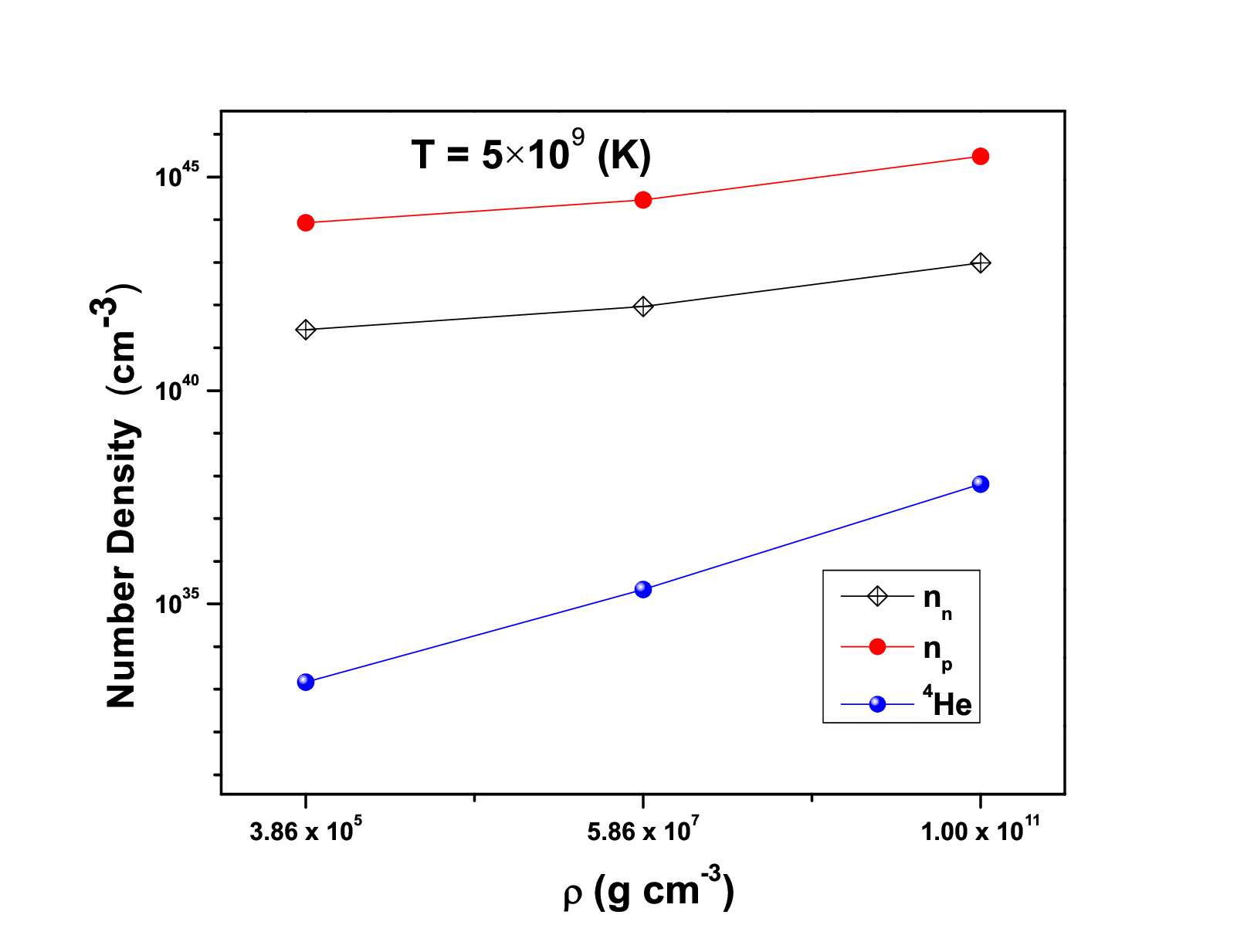}
\caption{(Color online) Number density of nucleons and alphas at
$T_{9} = 5$ as a function of stellar density.}\label{fig3}
\end{center}
\end{figure}

Figure~\ref{fig2} shows the number density of protons and neutrons
along with the production of alpha particles in stellar matter at
density $\rho_{st} = 10^{11}$ g/cm$^{3} $ as a function of stellar
temperature. It should be noted that $n_{n}$ and $n_{p}$ denote the
total number of neutrons and protons, respectively, present in our
ensemble.  One notes that the production of alphas increases with
stellar temperature. At low stellar temperatures the stellar matter
is mainly composed of heavy nuclei. With soaring temperatures heavy
nuclei gradually disintegrate into $\alpha$-particles and free
nucleons. Figure~\ref{fig3} presents a snapshot of number density of
nucleons and alphas at a fixed temperature of $T_9 = 5$. It is noted
that as the stellar core stiffens,  the number density of neutrons
and protons increases monotonically, albeit with a small slope. On
the other hand, the alpha particle production increases by orders of
magnitude mainly due to disintegration of heavy nuclei at $T_9 = 5$.

\begin{figure}
\begin{center}
\includegraphics[width=7.4in,height=6.6in]{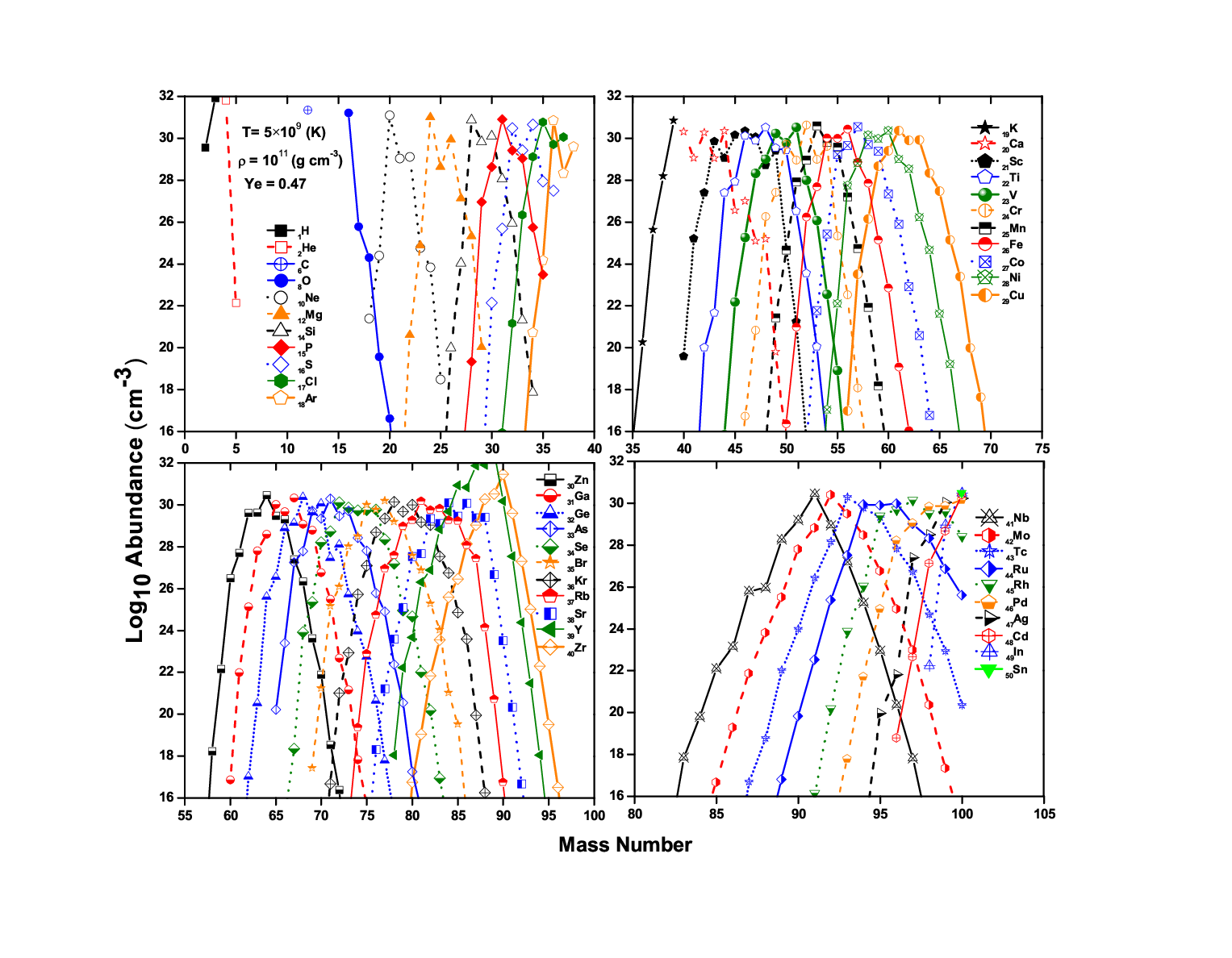}
\caption{(Color online)  Equilibrium abundances of nuclei at Fermi
energy of 23.91 MeV at $T_{9} = 5$.}\label{fig4}
\end{center}
\end{figure}

\begin{figure}
\begin{center}
\includegraphics[width=7.4in,height=6.6in]{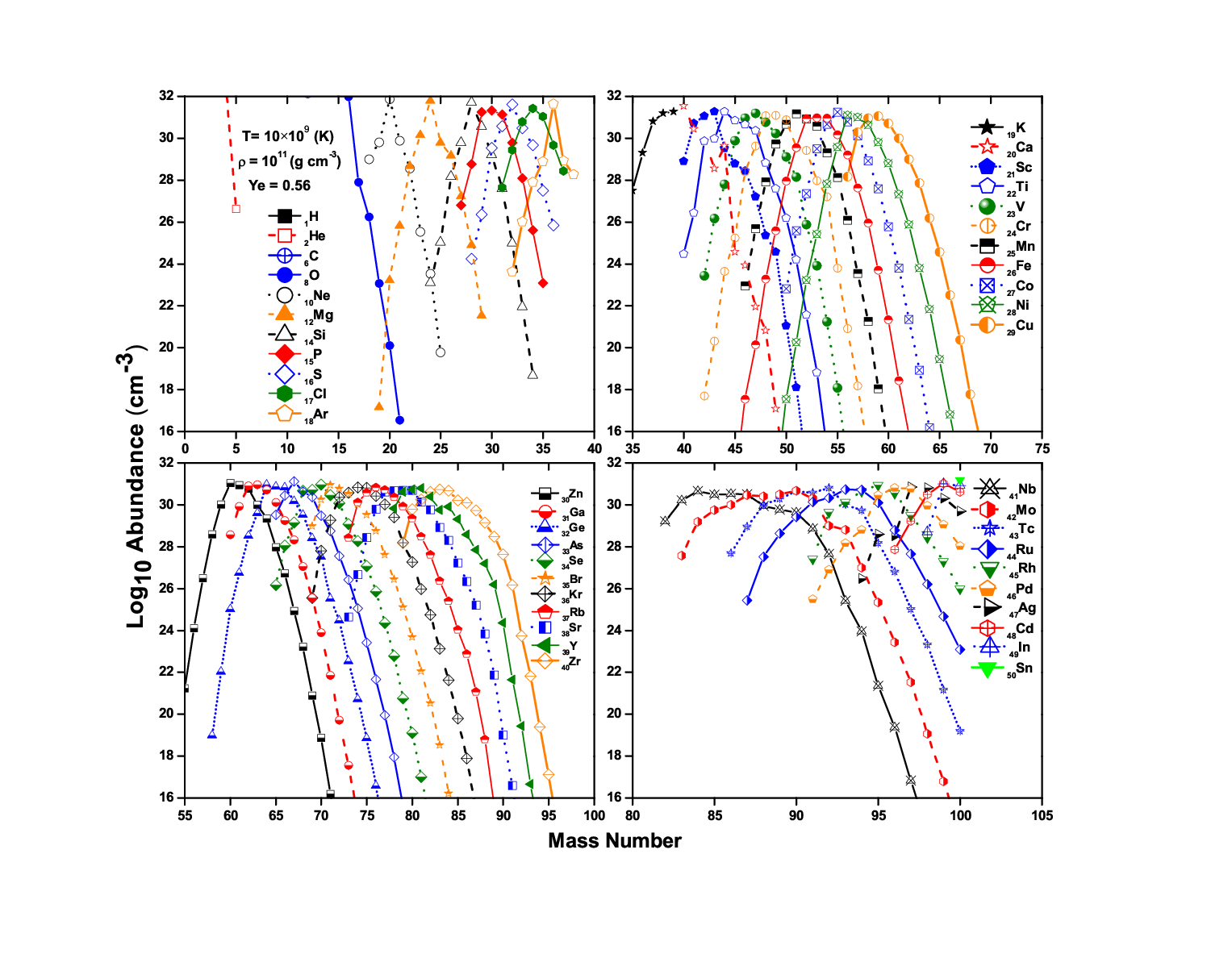}
\caption{(Color online)  Equilibrium abundances of nuclei at Fermi
energy of 23.83 MeV at $T_{9} = 10$.}\label{fig5}
\end{center}
\end{figure}

\begin{figure}
\begin{center}
\includegraphics[width=7.4in,height=6.6in]{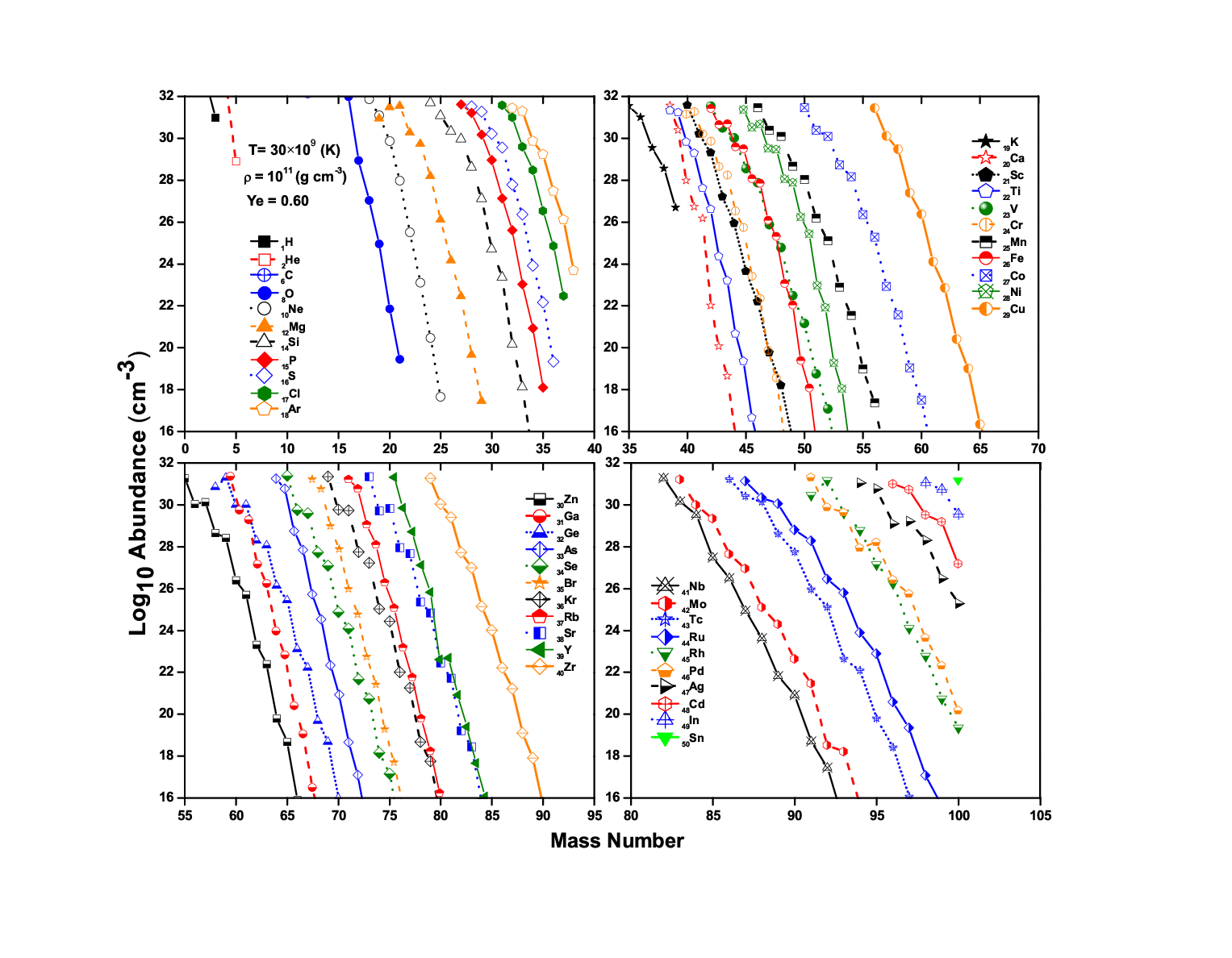}
\caption{(Color online)  Equilibrium abundances of nuclei at Fermi
energy of 23.02 MeV at $T_{9} = 30$.}\label{fig6}
\end{center}
\end{figure}

\begin{figure}
\begin{center}
\includegraphics[width=7.4in,height=6.6in]{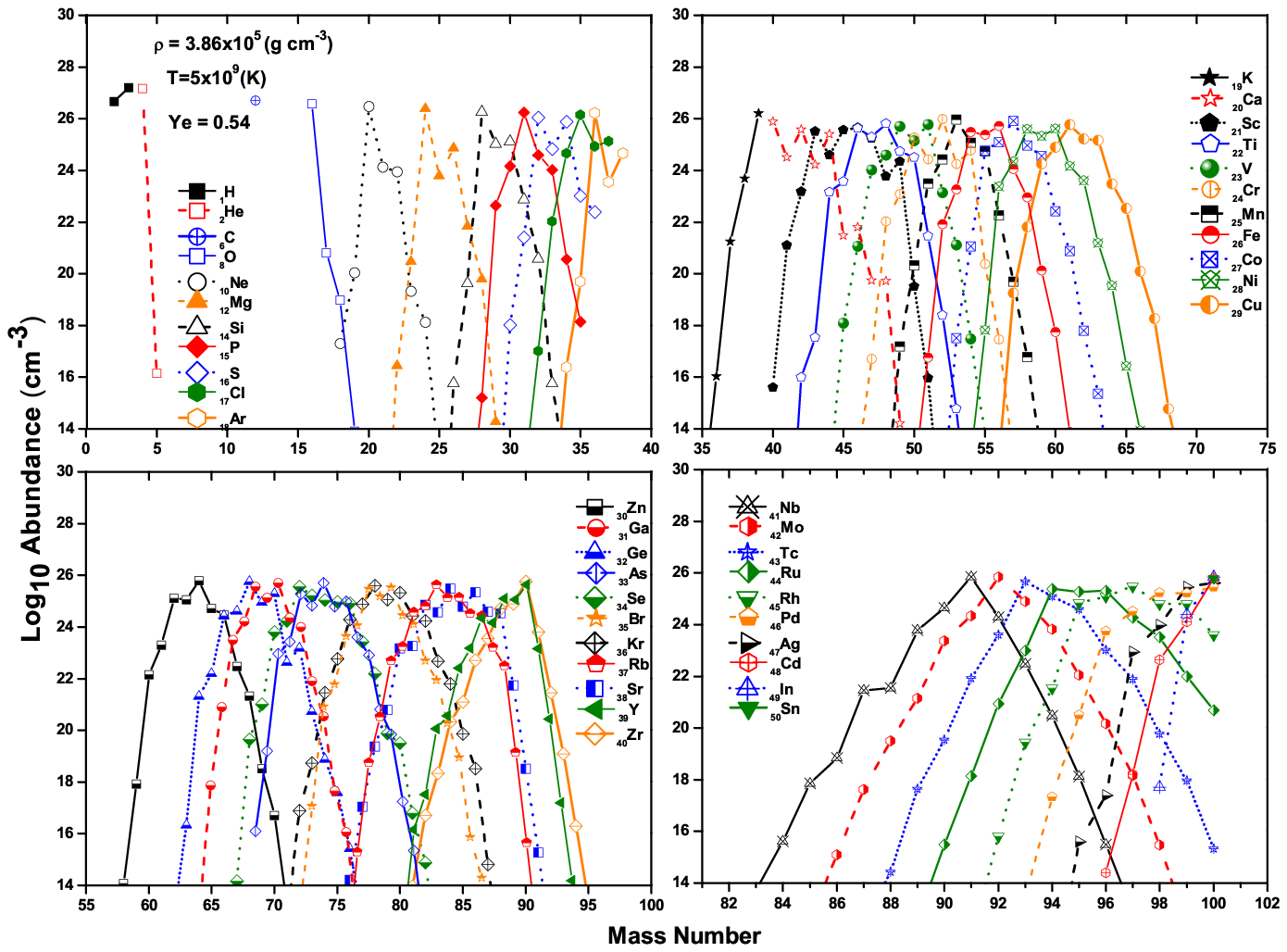}
\caption{(Color online)  Equilibrium abundances of nuclei at $T_{9}
= 5$ and $\rho_{st} =3.86 \times 10^{5}$ g/cm$^{3}$.}\label{fig7}
\end{center}
\end{figure}

\begin{figure}
\begin{center}
\includegraphics[width=7.4in,height=6.6in]{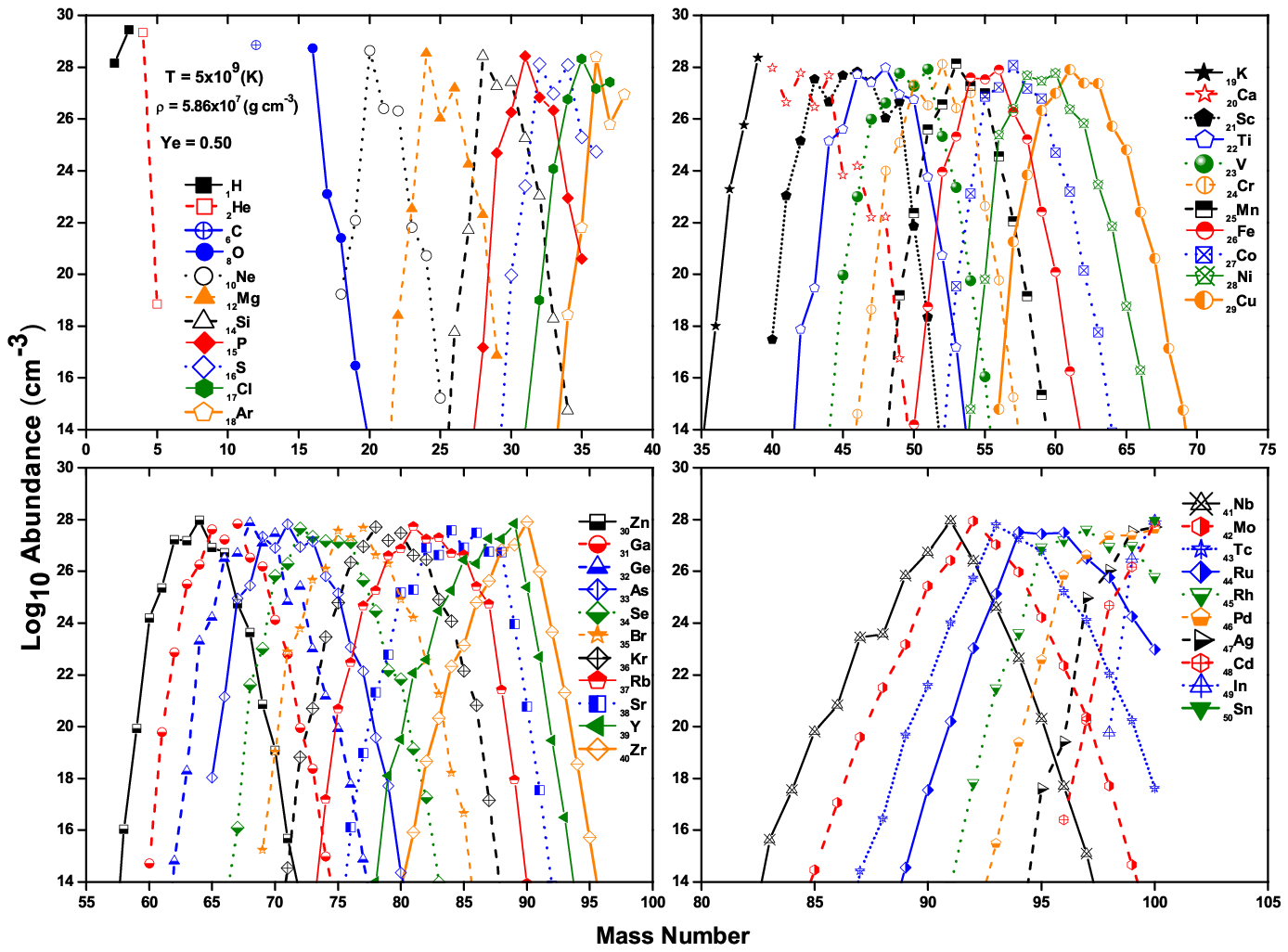}
\caption{(Color online)  Equilibrium abundances of nuclei at $T_{9}
= 5$ and $\rho_{st} =5.86 \times 10^{7}$ g/cm$^{3}$.}\label{fig8}
\end{center}
\end{figure}

Next we present our calculation of nuclear abundance assuming
conditions of NSE. Figures~\ref{fig4}-\ref{fig8} show the nuclear
abundance calculation (Eq.~\ref{na}) as function of stellar
temperatures and densities. Nuclear abundances are shown in
logarithmic scale (to base $10$) in all figures in units of
cm$^{-3}$. We first present our calculation of nuclear abundance as
a function of stellar temperature at fixed stellar densities. For
this purpose, we selected the  biggest ($\rho_{st} =10^{11}$
g/cm$^{3}$) density present in our calculation and show the results
of calculated nuclear abundance as a function of temperature. We
later investigate the equilibrium abundance pattern of the nuclei
present in our ensemble at a fixed temperature of $T_9 = 5$  as
function of core density.

Figures~\ref{fig4}-\ref{fig6} show the equilibrium abundance pattern
for nuclei at selected temperatures $T_9 = 5$, $T_9 = 10$  and $T_9
= 30$, respectively. In these figures the stellar density is held
fixed at $10^{11}$ g/cm$^{3} $. The equilibrium abundance pattern of
nuclei at $T_9 = 5$  is depicted in Figure~\ref{fig4}.
Neutron-number magic nuclei as well as doubly magic nuclei are
abundant for obvious reasons. Under prevailing physical conditions,
$^{3}$H and $^{4}$He are the most abundant nuclei possessing
abundances $\sim 10^{32}$ cm$^{-3}$. The least abundant nuclei are
$^{75}$Co and $^{78}$Ni with a negligible abundance of $ \sim
10^{-39}$ cm$^{-3}$.

Figure~\ref{fig5} shows calculated nuclear abundance at stellar
temperature $T_9 = 10$. Our calculation produced abundant quantity
of neutron-number magic nuclei number e.g $^{39}$K, $^{41}$Sc,
$^{56}$Ni, $^{62}$Zn including doubly magic nuclei $^{40}$Ca and
$^{100}$Sn. The most abundant nuclei in Figure~\ref{fig5} are again
$^{3}$H and $^{4}$He having abundance of the order of 10$^{33}$
cm$^{-3}$.

Figure~\ref{fig6} depicts the calculated nuclear abundance pattern
at temperature of $T_9 = 30$ and for $\rho_{st} = 10^{11}$
g/cm$^{3}$. We produce magic number nuclei e.g $^{40}$Ca,
$^{100}$In, $^{100}$Sn in good quantity.  $^{2}$H and $^{4}$He are
two most abundant nuclei under the given physical condition.
$^{78}$Ni is least abundant. We note that the abundance now is a
monotonic decreasing function of mass number.

We conclude from Figures~\ref{fig4}-\ref{fig6} that as temperature
increases the abundance distribution tends to become smooth as a
function of mass number. This is so mathematically because of the
exponential function in the fragment yield. Physically this is so
because temperature washes out the structure effects. Increasing
density does not have any significant impact on the abundance
pattern. Only the magnitude of nuclear abundance increases by orders
of magnitude. These abundance patterns contain key information about
fragmentation of matter in the nuclear liquid-gas phase transition.
At very low stellar densities the residual interaction between the
nuclide should be smallest. At high stellar temperatures the nuclear
matter disintegrates into small fragments and the nuclear abundances
decrease exponentially with mass number.

We next freeze the stellar temperature ($T_9 = 5$) and study the
equilibrium abundance pattern as a function of stellar density.
Figures~\ref{fig7}, \ref{fig8} and \ref{fig4}, respectively,
demonstrate our calculation of equilibrium abundance patterns for
successively higher densities. Figure~\ref{fig7} gives the abundance
pattern at a stellar density $\rho_{st} = 3.86 \times 10^{5}$
g/cm$^{3} $.  The abundance pattern increases with mass number,
shows a peak at neutron-number magic nuclei and then decreases with
increasing mass number. At the lowest Fermi energy, the equilibrium
configuration is peaked at $^{56}$Fe which is a well-known result
\cite{Tsu65} and reproduced by our calculation. As the stellar
density increases by another two order of magnitude to $\rho_{st} =
5.86 \times 10^{7}$ g/cm$^{3}$ (see Figure~\ref{fig8})  we note that
the equilibrium configuration remains almost unaltered but the
magnitude of the nuclear abundance increases roughly by the same
proportion as an increase in the density. When the stellar density
attains the maximum value of $\rho_{st} = 10^{11}$ g/cm$^{3} $ in
our calculation, we see from Figure~\ref{fig4} that the equilibrium
configuration remains more or less same with few orders of magnitude
increment in the corresponding nuclear abundances. Our results
obtained  maximum abundances for the nuclei $^{3}$H and $^{4}$He.
Our calculation also produced neutron magic number species  (e.g.
$^{38}$Ar, $^{39}$K, $^{40}$Ca, $^{54}$Fe, $^{55}$Co, $^{56}$Ni) at
above mentioned physical conditions in abundant quantities.

Our analysis shows that the equilibrium configuration of nuclei is
altered by changing the stellar temperatures. For a fixed stellar
temperature, increasing density by orders of magnitude has no effect
on the equilibrium configuration. Increasing stellar density only
increases the respective nuclear abundance magnitude but does not
have any effect on the distribution pattern.

\section{Summary and conclusions}
Conditions of NSE were assumed in order to perform nuclear abundance
calculations independent of specific supernova models for a total of
$728$ nuclei up to mass number $100$. The novelty of current
calculation lies in the treatment of TDNPFs. We performed a
state-by-state microscopic calculation of energy levels up to $10$
MeV in each nucleus. The energy eigenstates were calculated within
the pn-QRPA formalism. Experimental data was also inserted wherever
possible to further increase the reliability of the calculated
TDNPFs. Satisfactory convergence was achieved in our calculation of
the TDNPFs. The calculated partition functions were biggest for
odd-odd nuclei and smallest for even-even cases. The TDNPFs
increased with increasing temperature. The rate of change in the
TDNPFs increased with increasing stellar temperatures. Our
calculation of TDNPFs showed differences with the previous
calculations both at low and high stellar temperatures. For
even-even nuclei  the comparisons were in decent agreement.

The mass fractions were calculated using OSNSEC and the results were
in decent comparison to previous calculations \cite{Auf94, Liu07}.
Our calculation of nuclear abundances resulted in copious production
of few heavy nuclei, hydrogen-3, helium-4, carbon-12, oxygen-16 and
iron-group nuclei. At high stellar densities of the order $\rho_{st}
= 10^{10}$ g/cm$^{3} $ and beyond, $sd$-shell nuclei are the ones
with large abundance. The equilibrium abundance pattern was also
studied as function of the stellar temperatures and the densities.
Our study indicates that the equilibrium abundance pattern smooths
out with increasing temperature. As temperature increases so does
the production of isotopes with distinct peaks at magic number
nuclei. Further the abundance curve smooths out as the stellar
temperature rises. If we freeze the stellar temperature to a fixed
value and study the equilibrium abundance pattern as a function of
stellar density we note that there is no appreciable change in the
pattern. Only the abundance value increases by roughly the same
order of magnitude as the stellar density. ASCII files of TDNPFs (as
a function of stellar temperature) and nuclear abundances (as a
function of $\rho_{st}$, $T_{9}$ and $Y_{e}$) for all 728 nuclei may
be requested from the corresponding author.

It is hoped that this study will prove helpful in the nuclear
reaction network calculations. The results of this paper can prove
useful for the study of the presupernova stars (silicon burning
phases and beyond), the core-collapse/thermonuclear supernovae and
proto-neutron stars. We are in a process of integrating nuclear
abundance and weak interaction rates for all nuclei as function of
stellar densities and temperatures. This would help us to identify
key nuclei that should play a significant role in the stellar
evolution processes. Work is currently in progress and in future we
hope to publish list of top-$100$ weak-interaction nuclei that play
a crucial role in presupernova evolution of massive stars.

\textbf{Acknowledgements}

J.-U. Nabi wishes to acknowledge the support provided by the Higher
Education Commission (Pakistan) through the HEC Project No. 20-3099.
The work of A. Tawfik and N. Ezzelarab was supported by the World
Laboratory for Cosmology and Particle Physics (WLCAPP). The authors
are thankful to Adriana R. Raduta (IFIN-HH, Romania) for useful
comments on the manuscript.

\clearpage


\clearpage
\clearpage
\clearpage
\clearpage
\clearpage
\clearpage
\clearpage
\clearpage

\begin{center}
\scriptsize \footnotesize
\begin{longtable}{c|c|c|c }
\caption{ Table of nuclei considered in our ensemble of NSE.}\label{ta0}\\
 \hline
Nucleus&Mass Range&Nucleus&Mass Range\\
\hline
H&2-3&Cu&56-85\\
He&4-5&Zn&54-82\\
C&12-12&Ga&60-85\\
O&16-21&Ge&58-86\\
Ne&18-25&As&65-89\\
Mg&19-29&Se&65-92\\
Si&24-34&Br&69-94\\
P&27-35&Kr&69-97\\
S&28-36&Rb&73-100\\
Cl&31-37&Sr&73-100\\
Ar&32-38&Y&78-100\\
K&35-39&Zr&79-100\\
Ca&40-57&Nb&82-100\\
Sc&40-59&Mo&83-100\\
Ti&40-61&Tc&86-100\\
V&42-63&Ru&87-100\\
Cr&42-67&Rh&91-100\\
Mn&46-72&Pd&91-100\\
Fe&44-73&Ag&94-100\\
Co&50-79&Cd&96-100\\
Ni&49-78&In&98-100\\
Sn&100-100 & &\\
\hline
\end{longtable}
\end{center}
\clearpage

\begin{table}
\begin{center}
\caption {Comparison of calculated TDNPFs $(G)$ with those of Ref.
\cite{Rau00} for $^{46-67}$Fe. Numbers in parenthesis give stellar
temperature in units of 10$^{9}$ K.  } \tiny {
\begin{tabular}{c|c|c|c|c|c|c}
\hline
 \textbf{Fe} & \textbf{G(0.01,0.5,1)}  & \textbf{G(0.01,0.5,1)} & \textbf{G(2,4,5)}  & \textbf{G(2,4,5)} & \textbf{G(6,8,10)}  & \textbf{G(6,8,10)} \\
 & QRPA & FRDM & QRPA  & FRDM & QRPA  & FRDM \\
 \hline
    46 & 1.00E+00        & 1.00E+00 & 1.00E+00        & 1.00E+00 & 1.03E+00        & 1.14E+00 \\
     & 1.00E+00        & 1.00E+00 & 1.00E+00       & 1.01E+00 & 1.25E+00        & 1.63E+00 \\
     & 1.00E+00        & 1.00E+00 & 1.01E+00        & 1.05E+00 & 2.19E+00        & 3.08E+00 \\
     \hline
   47 & 8.00E+00        & 1.00E+00 & 8.15E+00        & 1.15E+00 & 9.17E+00       & 1.93E+00 \\
     & 8.00E+00       & 1.01E+00 & 8.61E+00        & 1.44E+00 & 1.03E+01        & 2.89E+00 \\
    & 8.02E+00        & 1.05E+00 & 8.87E+00        & 1.65E+00 & 1.39E+01        & 4.88E+00 \\
\end{tabular}}
\label{ta1}
\end{center}
\end{table}
\clearpage
\begin{table}
\caption {Comparison of calculated TDNPFs $(G)$ with those of Ref.
\cite{Rau03} for $^{46-67}$Fe. Numbers in parenthesis give stellar
temperature in units of 10$^{9}$ K.    } \tiny {
\begin{tabular}{c|c|c|c|c|c|c|c|c|c}
\hline \textbf{Fe}    & \textbf{G(12,14,16)} & \textbf{G(12,14,16)}
&\textbf{G(12,14,16)} & \textbf{G(18,20,22)} &
\textbf{G(18,20,22)} & \textbf{G(18,20,22)} &\textbf{G(24,26,28)} & \textbf{G(24,26,28)} & \textbf{G(24,26,28)} \\
   & QRPA & ETFSI &FRDM & QRPA & ETFSI & FRDM & QRPA & ETFSI & FRDM \\
\hline
46        & 5.07E+00 & 9.75E+00 & 6.80E+00 & 7.59E+01 & 9.57E+01 & 1.04E+02 & 6.35E+02 & 1.01E+03 & 1.48E+03 \\
          & 1.29E+01 & 2.04E+01 & 1.66E+01 & 1.67E+02 & 2.10E+02 & 2.57E+02 & 1.11E+03 & 2.21E+03 & 3.49E+03\\
          & 3.21E+01 & 4.38E+01 & 4.16E+01 & 3.38E+02 & 4.62E+02 & 6.22E+02 & 1.82E+03 & 4.81E+03 & 8.10E+03\\
          \hline
    47    & 2.62E+01 & 5.79E+00 & 9.04E+00 & 4.90E+02 & 4.38E+01 & 8.66E+01 & 4.97E+03 & 4.23E+02 & 9.61E+02\\
          & 6.64E+01 & 1.07E+01 & 1.84E+01 & 1.18E+03 & 9.23E+01 & 1.93E+02 & 8.94E+03 & 9.12E+02 & 2.14E+03\\
          & 1.85E+02 & 2.13E+01 & 3.94E+01 & 2.54E+03 & 1.97E+02 & 4.30E+02 & 1.50E+04 & 1.97E+03 & 4.76E+03\\
\end{tabular}}
\label{ta2}
\end{table}
\clearpage
\begin{table}
\begin{center}
\caption {Same as Tab.~\ref{ta1} but for $^{49-78}$Ni.  } \tiny {
\begin{tabular}{c|c|c|c|c|c|c|c|c|c}
\hline
 \textbf{Ni} & \textbf{G(0.01,0.5,1)} & \textbf{G(0.01,0.5,1)} & \textbf{G(0.01,0.5,1)} & \textbf{G(2,4,5)} &
  \textbf{G(2,4,5)} & \textbf{G(2,4,5)} & \textbf{G(6,8,10)} & \textbf{G(6,8,10)} & \textbf{G(6,8,10)} \\
   & QRPA & ETFSI & FRDM & QRPA & ETFSI & FRDM & QRPA & ETFSI & FRDM \\
 \hline
 49    & 1.00E+00 & 1.00E+00 & 1.00E+00 & 1.31E+00 & 1.02E+00 & 1.03E+00 & 2.31E+00 & 1.31E+00 & 1.43E+00 \\
          & 1.01E+00 & 1.00E+00 & 1.00E+00 & 1.77E+00 & 1.13E+00 & 1.17E+00 & 3.19E+00 & 1.63E+00 & 1.95E+00 \\
          & 1.08E+00 & 1.00E+00 & 1.00E+00 & 2.03E+00 & 1.21E+00 & 1.28E+00 & 6.30E+00 & 2.21E+00 & 3.04E+00 \\
          \hline
    50    & 1.00E+00 & 1.00E+00 & 1.00E+00 & 1.00E+00 & 1.00E+00 & 1.00E+00 & 1.03E+00 & 1.45E+00 & 1.12E+00 \\
          & 1.00E+00 & 1.00E+00 & 1.00E+00 & 1.00E+00 & 1.07E+00 & 1.01E+00 & 1.28E+00 & 2.48E+00 & 1.59E+00 \\
          & 1.00E+00 & 1.00E+00 & 1.00E+00 & 1.01E+00 & 1.20E+00 & 1.04E+00 & 2.37E+00 & 4.93E+00 & 3.06E+00 \\
\end{tabular}}
\label{ta3}
\end{center}
\end{table}
\clearpage
\begin{table}
\caption {Same as Tab.~\ref{ta2} but for $^{49-78}$Ni.  } \tiny {
\begin{tabular}{c|c|c|c|c|c|c|c|c|c}
\hline \textbf{Ni}    & \textbf{G(12,14,16)} & \textbf{G(12,14,16)}
&\textbf{G(12,14,16)} & \textbf{G(18,20,22)} &
 \textbf{G(18,20,22)} & \textbf{G(18,20,22)} &\textbf{G(24,26,28)} & \textbf{G(24,26,28)} & \textbf{G(24,26,28)} \\
   & QRPA & ETFSI &FRDM & QRPA & ETFSI & FRDM & QRPA & ETFSI & FRDM \\
\hline
    49    & 1.89E+01 & 3.06E+00 &-& 5.50E+02    & 1.75E+01    &- & 5.79E+03 & 1.61E+02 &-  \\
          & 6.33E+01 & 5.01E+00 &-& 1.35E+03    & 3.57E+01    &- & 1.05E+04 & 3.51E+02 & - \\
          & 1.98E+02 & 9.05E+00 &-& 2.94E+03    & 7.52E+01    &- & 1.76E+04 & 7.73E+02 &-  \\
          \hline
    50    & 5.84E+00 & 1.02E+01 & 6.95E+00 & 9.11E+01 & 1.22E+02 & 1.24E+02 & 7.51E+02 & 1.58E+03 & 2.10E+03 \\
          & 1.53E+01 & 2.28E+01 & 1.79E+01 & 1.99E+02 & 2.87E+02 & 3.24E+02 & 1.31E+03 & 3.71E+03 & 5.26E+03\\
          & 3.86E+01 & 5.24E+01 & 4.73E+01 & 4.02E+02 & 6.74E+02 & 8.31E+02 & 2.14E+03 & 8.67E+03 & 1.30E+04 \\
\end{tabular}}
\label{ta4}
\end{table}
\clearpage
\begin{table}
\begin{center}
\caption {Same as Tab.~\ref{ta1} but for $^{50-79}$Co.  } \tiny {
\begin{tabular}{c|c|c|c|c|c|c|c|c|c}
\hline
 \textbf{Co} & \textbf{G(0.01,0.5,1)} & \textbf{G(0.01,0.5,1)} & \textbf{G(0.01,0.5,1)} & \textbf{G(2,4,5)} &
  \textbf{G(2,4,5)} & \textbf{G(2,4,5)} & \textbf{G(6,8,10)} & \textbf{G(6,8,10)} & \textbf{G(6,8,10)} \\
   & QRPA & ETFSI & FRDM & QRPA & ETFSI & FRDM & QRPA & ETFSI & FRDM \\
 \hline
    50 & 1.30E+01 & 1.00E+00 & 1.00E+00 & 1.45E+01 & 1.15E+00 & 1.15E+00 & 1.89E+01 & 1.90E+00 & 2.07E+00\\
    & 1.31E+01 & 1.01E+00 & 1.01E+00 & 1.64E+01 & 1.43E+00 & 1.46E+00 & 2.30E+01 & 2.76E+00 & 3.30E+00 \\
    & 1.36E+01 & 1.05E+00 & 1.05E+00 & 1.75E+01 & 1.63E+00 & 1.72E+00 & 3.45E+01 & 4.45E+00 & 5.86E+00 \\
    \hline
    51  & 8.00E+00 & 1.00E+00 & 1.00E+00 & 8.56E+00 & 1.01E+00 & 1.00E+00 & 9.62E+00 & 1.25E+00 & 1.21E+00 \\
          & 8.05E+00 & 1.00E+00 & 1.00E+00 & 9.08E+00 & 1.08E+00 & 1.06E+00 & 1.10E+01 & 1.63E+00 & 1.58E+00 \\
          & 8.23E+00 & 1.00E+00 & 1.00E+00 & 9.32E+00 & 1.15E+00 & 1.12E+00 & 1.56E+01 & 2.46E+00 & 2.44E+00 \\
\end{tabular}}
\label{ta5}
\end{center}
\end{table}
\clearpage
\begin{table}
\caption {Same as Tab.~\ref{ta2} but for $^{50-79}$Co.  } \tiny {
\begin{tabular}{c|c|c|c|c|c|c|c|c|c}
\hline \textbf{Co}    & \textbf{G(12,14,16)} & \textbf{G(12,14,16)}
&\textbf{G(12,14,16)} & \textbf{G(18,20,22)} &
 \textbf{G(18,20,22)} & \textbf{G(18,20,22)} &\textbf{G(24,26,28)} & \textbf{G(24,26,28)} & \textbf{G(24,26,28)} \\
   & QRPA & ETFSI &FRDM & QRPA & ETFSI & FRDM & QRPA & ETFSI & FRDM \\
\hline
    50 & 7.52E+01 & 7.51E+00 & 1.09E+01 & 2.20E+03 & 5.75E+01 & 1.03E+02 & 2.64E+04 & 5.67E+02 & 1.17E+03 \\
      & 2.28E+02 & 1.41E+01 & 2.22E+01 & 5.70E+03 & 1.21E+02 & 2.29E+02 & 4.88E+04 & 1.24E+03 & 2.66E+03 \\
      & 7.40E+02 & 2.79E+01 & 4.71E+01 & 1.30E+04 & 2.61E+02 & 5.16E+02 & 8.34E+04 & 2.74E+03 & 6.08E+03 \\
      \hline
    51 & 3.10E+01 & 4.28E+00 & 4.41E+00 & 5.99E+02 & 4.32E+01 & 5.22E+01 & 6.11E+03 & 5.61E+02 & 7.69E+02 \\
          & 8.04E+01 & 8.62E+00 & 9.39E+00 & 1.44E+03 & 1.01E+02 & 1.28E+02 & 1.10E+04 & 1.33E+03 & 1.89E+03 \\
          & 2.25E+02 & 1.89E+01 & 2.17E+01 & 3.12E+03 & 2.38E+02 & 3.13E+02 & 1.84E+04 & 3.13E+03 & 4.63E+03 \\
\end{tabular}}
\label{ta6}
\end{table}
\clearpage
\begin{table}
\begin{center}
\caption {Same as Tab.~\ref{ta1} but for $^{56-85}$Cu.  } \tiny {
\begin{tabular}{c|c|c|c|c|c|c|c|c|c}
\hline
 \textbf{Cu} & \textbf{G(0.01,0.5,1)} & \textbf{G(0.01,0.5,1)} & \textbf{G(0.01,0.5,1)} & \textbf{G(2,4,5)} &
  \textbf{G(2,4,5)} & \textbf{G(2,4,5)} & \textbf{G(6,8,10)} & \textbf{G(6,8,10)} & \textbf{G(6,8,10)} \\
   & QRPA & ETFSI & FRDM & QRPA & ETFSI & FRDM & QRPA & ETFSI & FRDM \\
 \hline
 56       & 9.00E+00 & 1.00E+00 & 1.00E+00 & 9.95E+00 & 1.15E+00 & 1.46E+00 & 1.43E+01 & 1.93E+00 & 4.01E+00 \\
          & 9.14E+00 & 1.01E+00 & 1.04E+00 & 1.16E+01 & 1.44E+00 & 2.39E+00 & 1.96E+01 & 2.87E+00 & 7.12E+00 \\
          & 9.40E+00 & 1.05E+00 & 1.16E+00 & 1.27E+01 & 1.65E+00 & 3.08E+00 & 3.51E+01 & 4.81E+00 & 1.35E+01 \\
          \hline
    57    & 4.00E+00 & 1.00E+00 & 1.00E+00 & 4.48E+00 & 1.00E+00 & 1.00E+00 & 7.44E+00 & 1.37E+00 & 1.18E+00 \\
          & 4.00E+00 & 1.00E+00 & 1.00E+00 & 5.79E+00 & 1.09E+00 & 1.03E+00 & 1.01E+01 & 2.04E+00 & 1.61E+00 \\
          & 4.07E+00 & 1.00E+00 & 1.00E+00 & 6.57E+00 & 1.20E+00 & 1.09E+00 & 1.74E+01 & 3.60E+00 & 2.71E+00 \\
\end{tabular}}
\label{ta7}
\end{center}
\end{table}
\clearpage
\begin{table}
\caption {Same as Tab.~\ref{ta2} but for $^{56-85}$Cu.  } \tiny {
\begin{tabular}{c|c|c|c|c|c|c|c|c|c}
\hline \textbf{Cu}    & \textbf{G(12,14,16)} & \textbf{G(12,14,16)}
&\textbf{G(12,14,16)} & \textbf{G(18,20,22)} &
 \textbf{G(18,20,22)} & \textbf{G(18,20,22)} &\textbf{G(24,26,28)} & \textbf{G(24,26,28)} & \textbf{G(24,26,28)} \\
   & QRPA & ETFSI &FRDM & QRPA & ETFSI & FRDM & QRPA & ETFSI & FRDM \\
\hline
    56    & 9.53E+01 & 8.94E+00 &-& 3.49E+03 & 9.49E+01 &-& 4.23E+04 & 1.37E+03 &-  \\
          & 3.33E+02 & 1.85E+01 &- & 9.12E+03 & 2.27E+02 &-& 7.83E+04 & 3.42E+03 &-  \\
          & 1.15E+03 & 4.10E+01 &- & 2.08E+04 & 5.52E+02 &-& 1.34E+05 & 8.66E+03 & - \\
          \hline
    57    & 4.20E+01 & 7.04E+00 & 5.26E+00 & 9.42E+02 & 9.37E+01 & 8.06E+01 & 9.55E+03     & 1.60E+03 & 1.61E+03 \\
          & 1.22E+02 & 1.57E+01 & 1.22E+01 & 2.26E+03 & 2.38E+02 & 2.16E+02 & 1.72E+04    & 4.19E+03 & 4.43E+03 \\
          & 3.52E+02 & 3.77E+01 & 3.07E+01 & 4.88E+03 & 6.14E+02 &     5.86E+02 & 2.88E+04 & 1.11E+04 & 1.23E+04 \\
\end{tabular}}
\label{ta8}
\end{table}
\clearpage
\begin{table}
\begin{center}
\caption {Comparison of calculated TDNPFs $(G)$ with those used in
OSNSEC Ref. \cite{net} for $^{46-67}$Fe. Numbers in parenthesis give
stellar temperature in units of 10$^{9}$ K.  } \tiny {
\begin{tabular}{c|c|c|c|c|c|c}
    \hline
    \textbf{Fe} & \textbf{G(0.1,0.5,1)} & \textbf{G (0.1,0.5,1)} & \textbf{G(2.5,5,10)} & \textbf{G (2.5,5,10)} & \textbf{G(20,25,30)} & \textbf{G (20,25,30)} \\
             & QRPA  & OSNSEC  & QRPA     & OSNSEC & QRPA     & OSNSEC\\
    \hline
    46    & 1.00E+00 & 1.00E+00 & 1.00E+00 & 1.00E+00 & 1.67E+02 & 3.08E+00 \\
          & 1.00E+00 & 1.00E+00 & 1.01E+00 & 1.05E+00 & 8.47E+02 & 3.08E+00 \\
          & 1.00E+00 & 1.00E+00 & 2.19E+00 & 3.08E+00 & 2.84E+03 & 3.08E+00 \\
          \hline
    47    & 8.00E+00 & 6.00E+00 & 8.25E+00 & 7.27E+00 & 1.18E+03 & 2.93E+01 \\
          & 8.00E+00 & 6.06E+00 & 8.87E+00 & 9.92E+00 & 6.73E+03 & 2.93E+01 \\
          & 8.02E+00 & 6.30E+00 & 1.39E+01 & 2.93E+01 & 2.36E+04 & 2.93E+01 \\
\end{tabular}}
    \label{ta9}
\end{center}

\end{table}
\clearpage
\begin{table}
\begin{center}
\caption {Same as Tab.~\ref{ta9} but for $^{49-78}$Ni.} \tiny {
    \begin{tabular}{c|c|c|c|c|c|c}
    \hline
    \textbf{Ni} & \textbf{G(0.1,0.5,1)} & \textbf{G (0.1,0.5,1)} & \textbf{G(2.5,5,10)} & \textbf{G (2.5,5,10)} & \textbf{G(20,25,30)} & \textbf{G (20,25,30)} \\
             & QRPA  & OSNSEC  & QRPA     & OSNSEC & QRPA     & OSNSEC\\
    \hline
    49    & 1.00E+00 & 8.00E+00 & 1.42E+00 & 8.49E+00 & 1.35E+03 & 2.43E+01 \\
          & 1.01E+00 & 8.00E+00 & 2.03E+00 & 1.03E+01 & 7.86E+03 & 2.43E+01 \\
          & 1.08E+00 & 8.00E+00 & 6.30E+00 & 2.43E+01 & 2.78E+04 & 2.43E+01 \\
          \hline
    50    & 1.00E+00 & 1.00E+00 & 1.00E+00 & 1.00E+00 & 1.99E+02 & 3.06E+00 \\
          & 1.00E+00 & 1.00E+00 & 1.01E+00 & 1.04E+00 & 9.99E+02 & 3.06E+00 \\
          & 1.00E+00 & 1.00E+00 & 2.37E+00 & 3.06E+00 & 3.32E+03 & 3.06E+00 \\
\end{tabular}}
    \label{ta10}
\end{center}

\end{table}
\clearpage
\begin{table}
\begin{center}
\caption {Same as Tab.~\ref{ta9} but for $^{50-79}$Co.} \tiny {
    \begin{tabular}{c|c|c|c|c|c|c}
    \hline
    \textbf{Co} & \textbf{G(0.1,0.5,1)} & \textbf{G (0.1,0.5,1)} & \textbf{G(2.5,5,10)} & \textbf{G (2.5,5,10)} & \textbf{G(20,25,30)} & \textbf{G (20,25,30)} \\
             & QRPA  & OSNSEC  & QRPA     & OSNSEC & QRPA     & OSNSEC\\
    \hline
    50    & 1.30E+01 & 9.00E+00 & 1.50E+01 & 9.89E+00 & 5.70E+03 & 5.27E+01 \\
          & 1.31E+01 & 9.09E+00 & 1.75E+01 & 1.09E+01 & 3.63E+04 & 5.27E+01 \\
          & 1.36E+01 & 9.45E+00 & 3.45E+01 & 5.27E+01 & 1.33E+05 & 5.27E+01 \\
          \hline
    51    & 8.00E+00 & 8.00E+00 & 8.70E+00 & 8.00E+00 & 1.44E+03 & 1.95E+01 \\
          & 8.05E+00 & 8.00E+00 & 9.32E+00 & 8.08E+00 & 8.28E+03 & 1.95E+01 \\
          & 8.23E+00 & 8.00E+00 & 1.56E+01 & 1.95E+01 & 2.91E+04 & 1.95E+01 \\
\end{tabular}}
    \label{ta11}
\end{center}

\end{table}
\clearpage
\begin{table}
\begin{center}
\caption {Same as Tab.~\ref{ta9} but for $^{56-85}$Cu.} \tiny {
    \begin{tabular}{c|c|c|c|c|c|c}
    \hline
    \textbf{Cu} & \textbf{G(0.1,0.5,1)} & \textbf{G (0.1,0.5,1)} & \textbf{G(2.5,5,10)} & \textbf{G (2.5,5,10)} & \textbf{G(20,25,30)} & \textbf{G (20,25,30)} \\
             & QRPA  & OSNSEC  & QRPA     & OSNSEC & QRPA     & OSNSEC\\
    \hline
    56    & 9.00E+00 & 7.00E+00 & 1.03E+01 & 1.16E+01 & 9.12E+03 & 9.43E+01 \\
          & 9.14E+00 & 7.28E+00 & 1.27E+01 & 2.16E+01 & 5.82E+04 & 9.43E+01 \\
          & 9.40E+00 & 8.12E+00 & 3.51E+01 & 9.43E+01 & 2.14E+05 & 9.43E+01 \\
          \hline
    57    & 4.00E+00 & 4.00E+00 & 4.76E+00 & 4.00E+00 & 2.26E+03 & 1.08E+01 \\
          & 4.00E+00 & 4.00E+00 & 6.57E+00 & 4.36E+00 & 1.29E+04 & 1.08E+01 \\
          & 4.07E+00 & 4.00E+00 & 1.74E+01 & 1.08E+01 & 4.55E+04 & 1.08E+01 \\
\end{tabular}}
    \label{ta12}
\end{center}

\end{table}
\clearpage

\begin{table}
\caption{List of 20 nuclei with largest contribution to rate of
change of lepton fraction for $\rho_{st} =5.86 \times 10^{7}$
g/cm$^{3} $, $T_9 = 3.40$, $Y_e = 0.47$, along electron capture and
$\beta$-decay directions, as calculated by \cite{Auf94}. The list is
sorted according to largest mass fractions calculated by
\cite{Auf94}. Shown also is the corresponding mass fractions
calculated using the OSNSEC \cite{net}.} \label{ta13}
\begin{center}
\footnotesize    \begin{tabular}{ c | c | c | c |c |c }
    \multicolumn{3}{c|}{$\beta$-decay direction}&\multicolumn{3}{c}{electron capture direction}\\
    \hline
    $^A$Z&X$_k \cite{Auf94} $&X$_k $ (OSNSEC) &$^A$Z&X$_k \cite{Auf94}$&X$_k$ (OSNSEC)\\
    \hline
    $^{56}$Fe&5.05E-01&5.19E-01&$^{54}$Fe&2.11E-01&2.11E-01\\
    $^{55}$Fe&5.20E-02&5.50E-02&$^{60}$Ni&8.90E-02&7.60E-02\\
    $^{57}$Co&3.55E-02&3.10E-02&$^{58}$Ni&5.44E-02&5.43E-02\\
    $^{59}$Co&5.59E-04&4.21E-04&$^{55}$Fe&5.20E-02&5.50E-02\\
    $^{56}$Co&5.56E-04&9.02E-04&$^{57}$Co&3.55E-02&3.10E-02\\
    $^{55}$Mn&2.61E-04&3.54E-04&$^{53}$Mn&1.30E-02&1.39E-02\\
    $^{57}$Fe&2.60E-04&9.29E-04&$^{59}$Ni&5.13E-03&7.03E-03\\
    $^{54}$Mn&2.16E-04&5.27E-04&$^{56}$Co&5.56E-04&9.02E-04\\
    $^{58}$Co&1.31E-04&6.20E-04&$^{55}$Co&5.35E-04&5.60E-04\\
    $^{58}$Fe&8.28E-05&7.52E-05&$^{51}$Cr&3.57E-04&3.40E-04\\
    $^{53}$Cr&7.36E-05&6.14E-05&$^{54}$Mn&2.16E-04&5.27E-04\\
    $^{54}$Cr&4.58E-06&4.02E-06&$^{57}$Ni&1.71E-04&2.01E-04\\
    $^{50}$V&4.74E-07&7.39E-07&$^{58}$Co&1.31E-04&6.20E-04\\
    $^{60}$Co&1.02E-07&1.67E-07&$^{61}$Cu&5.00E-05&4.87E-05\\
    $^{56}$Mn&1.98E-08&5.06E-08&$^{52}$Mn&2.97E-05&3.93E-05\\
    $^{63}$Ni&5.78E-09&1.46E-08&$^{56}$Ni&2.72E-05&3.15E-05\\
    $^{64}$Cu&5.00E-09&1.12E-08&$^{53}$Fe&2.52E-05&2.83E-05\\
    $^{61}$Co&4.60E-09&2.87E-09&$^{62}$Cu&7.23E-07&2.76E-06\\
    $^{59}$Fe&2.06E-09&2.16E-09&$^{60}$Cu&3.07E-07&7.69E-07\\
    $^{52}$V&4.32E-10&1.23E-09&$^{59}$Cu&2.69E-07&3.12E-07\\
    \hline
    \end{tabular}
\end{center}
\end{table}
\clearpage
\begin{table}
\begin{center}
\caption{Top 30 abundant nuclei according to our calculation for
physical conditions mentioned in Tab.~\ref{ta13}. } \label{ta14}
    \begin{tabular}{c|c|r}
    Elements & A     & X(A,Z) \\
\hline
    Fe &56  & 5.19E-01 \\
    Fe &54  & 2.11E-01 \\
    Ni &60  & 7.60E-02 \\
    Fe &55  & 5.50E-02 \\
    Ni &58  & 5.43E-02 \\
    Co &57  & 3.10E-02 \\
    Cr &52  & 2.57E-02 \\
    Mn &53  & 1.39E-02 \\
    Ni &59  & 7.03E-03 \\
    Cr &50  & 1.07E-03 \\
    Fe &57  & 9.29E-04 \\
    Co &56  & 9.02E-04 \\
    Co &58  & 6.20E-04 \\
    Co &55  & 5.60E-04 \\
    Mn &54  & 5.27E-04 \\
    Co &59  & 4.21E-04 \\
    Mn &55  & 3.54E-04 \\
    Cr &51  & 3.40E-04 \\
    Ni &61  & 2.73E-04 \\
    Ni &57  & 2.01E-04 \\
    Ni &62  & 1.29E-04 \\
    Fe &58  & 7.52E-05 \\
    Cr &53  & 6.14E-05 \\
    Cu &61  & 4.87E-05 \\
    Mn &52  & 3.93E-05 \\
    Ni &56  & 3.15E-05 \\
    Fe &53  & 2.83E-05 \\
    V  &49   & 7.02E-06 \\
    Cu &63  & 6.96E-06 \\
    V  &51   & 6.28E-06 \\
\hline
\end{tabular}
\end{center}
\end{table}
\clearpage
\begin{table}
\caption{Same as Tab.~\ref{ta13} but for $\rho_{st} =1.45 \times
10^{8}$ g/cm$^{3} $, $T_9 = 3.80$,
    $Y_e = 0.45$.} \label{ta15}
\begin{center}
\footnotesize    \begin{tabular}{ c | c | c | c |c |c }
    \multicolumn{3}{c|}{$\beta$-decay direction}&\multicolumn{3}{c}{electron capture direction}\\
    \hline
    $^A$Z&X$_k \cite{Auf94} $&X$_k $ (OSNSEC) &$^A$Z&X$_k \cite{Auf94}$&X$_k$ (OSNSEC)\\
    \hline
    $^{58}$Fe&4.17E-01&4.31E-01&$^{56}$Fe&8.30E-02&7.84E-02\\
    $^{54}$Cr&1.63E-01&1.69E-01&$^{55}$Mn&2.94E-02&3.96E-02\\
    $^{53}$Cr&2.06E-02&1.77E-02&$^{53}$Cr&2.06E-02&1.77E-02\\
    $^{57}$Fe&1.19E-02&4.10E-02&$^{57}$Fe&1.19E-02&4.10E-02\\
    $^{59}$Fe&4.17E-03&5.42E-03&$^{59}$Co&1.14E-02&8.14E-03\\
    $^{61}$Co&3.95E-03&2.83E-03&$^{51}$V&5.93E-03&5.71E-03\\
    $^{63}$Ni&1.92E-03&5.51E-03&$^{60}$Ni&3.36E-03&2.42E-03\\
    $^{60}$Fe&1.90E-03&1.97E-03&$^{61}$Ni&1.46E-03&2.50E-03\\
    $^{56}$Mn&7.03E-04&2.02E-03&$^{56}$Mn&7.03E-04&2.02E-03\\
    $^{57}$Mn&6.72E-04&8.65E-04&$^{60}$Co&6.03E-04&1.08E-03\\
    $^{60}$Co&6.03E-04&1.08E-03&$^{54}$Mn&2.30E-04&5.20E-04\\
    $^{55}$Cr&5.85E-04&7.43E-04&$^{55}$Fe&1.14E-04&9.80E-05\\
    $^{51}$Ti&1.28E-04&1.01E-04&$^{53}$Mn&8.08E-05&7.00E-05\\
    $^{52}$V&1.17E-04&3.80E-04&$^{63}$Cu&5.16E-05&3.67E-05\\
   $^{53}$V&9.45E-05&1.25E-04&$^{57}$Co&3.84E-05&2.56E-05\\
   $^{65}$Ni&6.61E-05&7.44E-05&$^{58}$Co&2.79E-05&1.13E-04\\
   $^{62}$Co&7.41E-06&1.83E-05&$^{51}$Cr&7.44E-06&5.89E-06\\
   $^{63}$Co&6.93E-06&5.26E-06&$^{64}$Cu&7.37E-06&1.73E-05\\
   $^{61}$Fe&1.31E-06&9.21E-07&$^{50}$V&5.32E-06&8.02E-06\\
   $^{58}$Mn&1.20E-06&1.92E-06&$^{62}$Cu&5.08E-08&1.66E-07\\
    \hline
    \end{tabular}
\end{center}
\end{table}
\clearpage
\begin{table}
\begin{center}
\caption{Top 30 abundant nuclei according to our calculation for
physical conditions mentioned in Tab.~\ref{ta15}. } \label{ta16}
    \begin{tabular}{c|c|r}
    Elements & A     & X(A,Z) \\
\hline
    Fe &58  & 4.31E-01 \\
    Cr &54  & 1.69E-01 \\
    Ni &62  & 1.20E-01 \\
    Fe &56  & 7.84E-02 \\
    Fe &57  & 4.10E-02 \\
    Mn &55  & 3.96E-02 \\
    Cr &52  & 2.69E-02 \\
    Ti &50  & 2.08E-02 \\
    Cr &53  & 1.77E-02 \\
    Ni &64  & 1.51E-02 \\
    Co &59  & 8.14E-03 \\
    V  &51   & 5.71E-03 \\
    Ni &63  & 5.51E-03 \\
    Fe &59  & 5.42E-03 \\
    Co &61  & 2.83E-03 \\
    Ni &61  & 2.50E-03 \\
    Ni &60  & 2.42E-03 \\
    Mn &56  & 2.02E-03 \\
    Fe &60  & 1.97E-03 \\
    Co &60  & 1.08E-03 \\
    Mn &57  & 8.65E-04 \\
    Cr &55  & 7.43E-04 \\
    Mn &54  & 5.20E-04 \\
    V  &52   & 3.80E-04 \\
    Cu &65  & 2.66E-04 \\
    V  &53   & 1.25E-04 \\
    Ti &49  & 1.21E-04 \\
    Co &58  & 1.13E-04 \\
    Ti &51  & 1.01E-04 \\
    Fe &55  & 9.80E-05 \\
\hline
\end{tabular}
\end{center}
\end{table}
\clearpage
\begin{table}
\caption{Same as Tab.~\ref{ta13} but for $\rho_{st} =1.06 \times
10^{9}$ g/cm$^{3} $, $T_9 = 4.93$,
    $Y_e = 0.43$.} \label{ta17}
\begin{center}
\footnotesize    \begin{tabular}{ c | c | c | c |c |c }
    \multicolumn{3}{c|}{$\beta$-decay direction}&\multicolumn{3}{c}{electron capture direction}\\
    \hline
    $^A$Z&X$_k \cite{Auf94} $&X$_k $ (OSNSEC) &$^A$Z&X$_k \cite{Auf94}$&X$_k$ (OSNSEC)\\
    \hline
    $^{60}$Fe&1.00E-01&1.24E-01&$^{59}$Fe&1.46E-02&2.00E-02\\
    $^{65}$Ni&3.87E-02&4.96E-02&$^{57}$Mn&8.95E-03&1.15E-02\\
    $^{51}$Ti&2.91E-02&2.45E-02&$^{61}$Co&6.60E-03&4.47E-03\\
    $^{67}$Ni&2.11E-02&2.35E-02&$^{67}$Cu&6.34E-03&6.72E-03\\
    $^{68}$Ni&2.09E-02&1.98E-02&$^{58}$Mn&1.44E-03&2.63E-03\\
    $^{55}$Cr&1.78E-02&2.51E-02&$^{63}$Ni&1.33E-03&3.35E-03\\
    $^{63}$Co&1.69E-02&1.56E-02&$^{62}$Co&1.03E-03&2.80E-03\\
    $^{59}$Fe&1.46E-02&2.00E-02&$^{68}$Cu&8.35E-04&2.19E-03\\
    $^{61}$Fe&1.06E-02&9.86E-03&$^{51}$V&8.05E-04&6.54E-04\\
    $^{62}$Fe&8.38E-03&1.23E-02&$^{52}$V&7.40E-04&2.22E-03\\
    $^{57}$Cr&1.91E-03&2.12E-03&$^{56}$Mn&5.62E-04&1.47E-03\\
    $^{59}$Mn&1.76E-03&3.36E-03&$^{55}$Mn&4.85E-04&5.16E-04\\
    $^{50}$Sc&1.74E-03&2.41E-03&$^{65}$Cu&1.84E-04&1.11E-04\\
    $^{58}$Mn&1.44E-03&2.63E-03&$^{66}$Cu&1.31E-04&2.68E-04\\
    $^{62}$Co&1.03E-03&2.80E-03&$^{49}$Ti&1.18E-04&7.76E-05\\
    $^{65}$Co&9.56E-04&1.69E-03&$^{60}$Co&9.51E-05&1.57E-04\\
    $^{54}$V&8.33E-04&1.27E-03&$^{48}$Sc&5.93E-05&9.76E-05\\
    $^{64}$Co&5.68E-04&1.58E-03&$^{59}$Co&4.42E-05&2.41E-05\\
    $^{58}$Cr&4.99E-04&3.53E-04&$^{61}$Ni&3.41E-06&4.42E-06\\
    $^{63}$Fe&5.10E-05&2.31E-04&$^{64}$Cu&5.41E-07&1.06E-06\\
    \hline
    \end{tabular}
\end{center}
\end{table}
\clearpage
\begin{table}
\begin{center}
\caption{Top 30 abundant nuclei according to our calculation for
physical conditions mentioned in Tab.~\ref{ta17}. } \label{ta18}
    \begin{tabular}{c|c|r}
    Elements & A     & X(A,Z) \\
\hline
    Ni &66  & 2.34E-01 \\
    Fe &60  & 1.24E-01 \\
    Ni &64  & 8.34E-02 \\
    Cr &54  & 5.51E-02 \\
    Ti &50  & 5.19E-02 \\
    Ni &65  & 4.96E-02 \\
    Ca &48  & 4.80E-02 \\
    Cr &56  & 3.89E-02 \\
    Ti &52  & 2.67E-02 \\
    Cr &55  & 2.51E-02 \\
    Ti &51  & 2.45E-02 \\
    Ni &67  & 2.35E-02 \\
    Fe &59  & 2.00E-02 \\
    Ni &68  & 1.98E-02 \\
    Fe &58  & 1.98E-02 \\
    Sc &49  & 1.89E-02 \\
    V  &53  & 1.61E-02 \\
    Co &63  & 1.56E-02 \\
    Fe &62  & 1.23E-02 \\
    Mn &57  & 1.15E-02 \\
    Fe &61  & 9.86E-03 \\
    Cu &67  & 6.72E-03 \\
    Co &61  & 4.47E-03 \\
    Cu &69  & 4.19E-03 \\
    Zn &72  & 3.71E-03 \\
    Mn &59  & 3.36E-03 \\
    Ni &63  & 3.35E-03 \\
    Co &62  & 2.80E-03 \\
    Mn &58  & 2.63E-03 \\
    Zn &70  & 2.60E-03 \\
\hline
\end{tabular}
\end{center}
\end{table}
\clearpage
\begin{table}
\caption{Same as Tab.~\ref{ta13} but for $\rho_{st} =4.01 \times
10^{10}$ g/cm$^{3} $, $T_9 = 7.33$,
    $Y_e = 0.41$.} \label{ta19}
\begin{center}
\footnotesize    \begin{tabular}{ c | c | c | c |c |c }
    \multicolumn{3}{c|}{$\beta$-decay direction}&\multicolumn{3}{c}{electron capture direction}\\
    \hline
    $^A$Z&X$_k \cite{Auf94} $&X$_k $ (OSNSEC) &$^A$Z&X$_k \cite{Auf94}$&X$_k$ (OSNSEC)\\
    \hline
    $^{49}$Ca&3.88E-02&2.07E-02&$^{49}$Sc&1.39E-02&8.38E-03\\
    $^{53}$Ti&1.28E-02&1.20E-02&$^{50}$Sc&1.03E-02&1.13E-02\\
    $^{51}$Sc&1.08E-02&1.17E-02&$^{61}$Fe&6.09E-03&6.98E-03\\
    $^{50}$Sc&1.03E-02&1.13E-02&$^{57}$Cr&5.92E-03&7.75E-03\\
    $^{50}$Ca&7.68E-03&1.27E-02&$^{55}$V&5.46E-03&7.51E-03\\
    $^{55}$V&5.46E-03&3.95E-04&$^{56}$Cr&5.34E-03&8.94E-03\\
    $^{66}$Co&4.42E-03&6.58E-03&$^{63}$Co&4.78E-03&5.97E-03\\
    $^{54}$Ti&4.17E-03&5.00E-03&$^{65}$Ni&4.43E-03&6.26E-03\\
    $^{61}$Mn&2.65E-03&2.57E-03&$^{60}$Fe&4.22E-03&6.20E-03\\
    $^{63}$Fe&2.42E-03&7.58E-03&$^{59}$Mn&3.16E-03&6.70E-03\\
    $^{60}$Mn&1.49E-03&4.51E-03&$^{53}$V&2.76E-03&4.11E-03\\
    $^{52}$Sc&1.24E-03&2.46E-03&$^{64}$Co&2.56E-03&6.64E-03\\
    $^{59}$Cr&8.58E-04&1.03E-03&$^{54}$V&2.15E-03&3.68E-03\\
    $^{68}$Co&8.07E-04&6.29E-04&$^{60}$Mn&1.49E-03&4.51E-03\\
    $^{53}$Sc&5.70E-04&1.55E-04&$^{68}$Cu&1.19E-03&3.21E-03\\
    $^{56}$V&5.16E-04&1.24E-03&$^{58}$Mn&9.84E-04&2.02E-03\\
    $^{55}$Ti&4.97E-04&4.24E-04&$^{57}$Mn&6.87E-04&1.15E-03\\
    $^{57}$V&3.77E-04&3.95E-04&$^{59}$Fe&5.06E-04&9.15E-04\\
    $^{51}$Ca&3.13E-04&5.99E-04&$^{62}$Co&2.45E-04&8.14E-04\\
    $^{62}$Mn&1.53E-04&2.35E-04&$^{61}$Co&1.80E-04&1.99E-04\\
    \hline
    \end{tabular}
\end{center}
\end{table}
\clearpage
\begin{table}
\begin{center}
\caption{Top 30 abundant nuclei according to our calculation for
physical conditions mentioned in Tab.~\ref{ta19}. } \label{ta20}
    \begin{tabular}{c|c|r}
    Elements & A     & X(A,Z) \\
\hline
    Ge &80  & 5.63E-02 \\
    Ni &68  & 5.13E-02 \\
    Ca &48  & 4.48E-02 \\
    As &83  & 3.91E-02 \\
    Ni &67  & 3.55E-02 \\
    Se &84  & 3.03E-02 \\
    Ni &66  & 2.84E-02 \\
    Ga &77  & 2.80E-02 \\
    Ge &79  & 2.63E-02 \\
    As &81  & 2.59E-02 \\
    As &82  & 2.38E-02 \\
    Ge &78  & 2.32E-02 \\
    Cu &71  & 2.12E-02 \\
    Ge &81  & 2.12E-02 \\
    Ca &49  & 2.07E-02 \\
    Ni &69  & 2.05E-02 \\
    Ti &52  & 2.05E-02 \\
    Zn &74  & 1.98E-02 \\
    Se &85  & 1.92E-02 \\
    Se &83  & 1.67E-02 \\
    Co &65  & 1.61E-02 \\
    Fe &62  & 1.40E-02 \\
    Zn &73  & 1.39E-02 \\
    Ca &50  & 1.27E-02 \\
    Cu &70  & 1.25E-02 \\
    Zn &76  & 1.24E-02 \\
    Ti &53  & 1.20E-02 \\
    Ni &70  & 1.20E-02 \\
    Sc &51  & 1.17E-02 \\
    Sc &50  & 1.13E-02 \\
\hline
\end{tabular}
\end{center}
\end{table}
\clearpage
\begin{table}
\caption{Mass fractions for $^{52,53,54}$Fe, $^{55}$Co and
$^{56,58}$Ni at temperature $T= 5\times10^{9}$K, $\rho_{st} =
1\times10^{7}$ g.cm$^{-3}$ and $Y_{e}= 0.5$ calculated using the
OSNSEC \cite{net} compared with the calculation of \cite{Liu07}.}
\label{ta21}
\begin{center}
\footnotesize\begin{tabular}{c|c|c} \hline
    Nuclei& $X(A,Z)$ (OSNSEC) & $X(A,Z)$\cite{Liu07}\\
    \hline

$^{52}$Fe& 3.11$\times10^{-2}$ &2.58$\times10^{-2}$\\
$^{53}$Fe&2.53$\times10^{-2}$  &3.65$\times10^{-2}$\\
$^{54}$Fe&2.26$\times10^{-1}$ &2.98$\times10^{-1}$\\
$^{55}$Co&1.55$\times10^{-1}$ &2.87$\times10^{-2}$\\
$^{56}$Ni&2.70$\times10^{-1}$  &3.98$\times10^{-1}$\\
$^{58}$Ni&8.10$\times10^{-2}$ &1.07$\times10^{-1}$\\
\hline
\end{tabular}
\end{center}
\end{table}
\end{document}